\documentclass[twocolumn,showpacs,preprintnumbers,amsmath,amssymb]{revtex4}
\usepackage{graphicx,epsfig}
\usepackage{dcolumn}
\usepackage{bm}
\newcommand{\bi}{\begin{itemize}}
\newcommand{\ei}{\end{itemize}}
\newcommand{\be}{\begin{equation}}
\newcommand{\ee}{\end{equation}}
\newcommand{\ba}{\begin{eqnarray}}
\newcommand{\ea}{\end{eqnarray}}
\newcommand{\bse}{\begin{subequations}}
\newcommand{\ese}{\end{subequations}}

\newcommand{\C}{{\cal {C}}}

\newcommand{\bben}{\begin{itemize}}
\newcommand{\eeen}{\end{itemize}}
\newcommand{\bbq}{\begin{quote}}
\newcommand{\eeq}{\end{quote}}

\newcommand{\tbb}{t_{\textrm{\tiny{bb}}}}
\newcommand{\taub}{\tau_{\textrm{\tiny{bb}}}}
\newcommand{\RR}{{}^3{\cal{R}}}
\newcommand{\RRi}{{}^3{\cal{R}}_i}
\newcommand{\T}{{}^3{\cal{T}}}
\newcommand{\Ti}{{}^3{\cal{T}}_i}
\newcommand{\EE}{{\cal{E}}}
\newcommand{\FF}{{\cal{F}}}
\newcommand{\VV}{{\cal{V}}}
\newcommand{\HH}{{\cal{H}}}

\newcommand{\rhoiav}{\langle\rho_i\rangle}
\newcommand{\miav}{\langle m_i\rangle}
\newcommand{\Omiav}{\langle\Omega_i\rangle}
\newcommand{\Omiavr}{\langle\Omega_i(r)\rangle}
\newcommand{\RRiav}{\langle\RRi\rangle}
\newcommand{\kiav}{\langle k_i \rangle}
\newcommand{\rhoav}{\langle\rho\rangle}
\newcommand{\mav}{\langle m\rangle}
\newcommand{\RRav}{\langle\RR\rangle}
\newcommand{\kav}{\langle k \rangle}
\newcommand{\Omav}{\langle\Omega\rangle}
\newcommand{\Omavxr}{\langle\Omega(\xi,r)\rangle}

\newcommand{\Dim}{\Delta_i^{(m)}}
\newcommand{\Dik}{\Delta_i^{(k)}}
\newcommand{\Dm}{\Delta^{(m)}}
\newcommand{\Dk}{\Delta^{(k)}}

\begin{document}

\title{A dynamical system approach to inhomogeneous dust solutions.} 

\author{ 
Roberto A. Sussman$^\ddagger$}

\email{sussman@nucleares.unam.mx}

\affiliation{ Instituto de F\'\i sica, Universidad de Guanajuato, Loma del Bosque 103, Leon, Guanajuato, 37150, M\'exico. 
 $^\ddagger$ On sabbatical leave from Instituto de Ciencias Nucleares, Universidad Nacional  Aut\'onoma de M\'exico (ICN-UNAM), A. P. 70--543, 04510 M\'exico D. F., 
M\'exico.}


\begin{abstract}
We examine numerically and qualitatively the Lema\^\i tre--Tolman--Bondi (LTB) inhomogeneous dust solutions as a 3--dimensional dynamical system characterized by six critical points. One of the coordinates of the phase space is an average  density parameter, $\Omav$, which behaves as the ordinary $\Omega$ in Friedman-Lema\^\i tre--Robertson--Walker (FLRW) dust spacetimes. The other two coordinates, a shear parameter and a density contrast function, convey the effects of inhomogeneity. As long as shell crossing singularities are absent, this phase space is bounded or it can be trivially compactified. This space contains several invariant subspaces which define relevant particular cases, such as:  ``parabolic'' evolution, FLRW dust and the Schwarzschild--Kruskal vacuum limit.   We examine in detail the phase space evolution of several dust configurations: a low density void formation scenario, high density re--collapsing universes with open, closed and wormhole topologies, a structure formation scenario with a black hole surrounded by an expanding background, and the Schwarzschild--Kruskal vacuum case.  Solution curves (except regular centers) start expanding from a past attractor (source) in the plane $\Omav=1$, associated with self similar regime at an initial singularity. Depending on the initial conditions and specific configurations, the curves approach several saddle points as they evolve between this past attractor and other two possible future attractors: perpetually expanding curves terminate at a line of sinks at $\Omav=0$, while collapsing curves reach maximal expansion as $\Omav$ diverges and end up in sink that coincides with the past attractor and is also associated with self similar behavior.     

\end{abstract}


\maketitle

\section{Introduction} 

Inhomogeneous dust solutions with spherical symmetry are among the oldest, simplest and most useful exact solutions of Einstein's equations. These solutions were initially derived independently by Lema\^\i tre (1933) and Tolman (1934) and then re-derived by Bondi (1947), hence they are known as the Lema\^\i tre--Tolman--Bondi (LTB) solutions (see \cite{kras} and references quoted therein for a comprehensive review on these solutions and their applications). 

In practically all applications of these solutions the standard original variables in which they were derived are used, though alternative variables amenable to an ``initial value'' treatment have been proposed (see \cite{SG} and references quoted therein). While not the only type of parametrization for these solutions, we find these variables particularly useful for a numerical treatment. 

We propose in this paper to examine these models within the framework of numeric and qualitative techniques known generically as ``dynamical systems''~\cite{EW}. This approach requires as a first necessary step expressing the evolution equations as a proper system of first order autonomous ODE's (ordinary differential equations). This necessary requirement would, apparently, exclude inhomogeneous LTB solutions because their evolution equations are necessarily PDE's (partial differential equations). However, under a ``3+1'' covariant decomposition based on the 4--velocity field~\cite{EVE}, the evolution equations for LTB dust solutions become first order autonomous PDE's containing only time derivatives~\cite{EW}. We prove in this case how the spacelike gradient equations (constraints) are automatically satisfied for all times once initial conditions are selected so that these constraints hold in an arbitrary initial hypersurface $\T_i$ of constant time. Rigorously (see Appendix B), the solution curves of these evolution equations are then equivalent to a proper dynamical system built with ODE's, under the strong restriction of complying with the special set of initial conditions that is compatible with the spacelike constraints of the PDE system.

In order to describe LTB dust solutions as a dynamical system we re--write the first order evolution equations of the 3+1 decomposition~\cite{EVE} in terms of initial value variables defined in reference \cite{SG}. These variables are introduced and generalized in section III and in sections IV and V we show how they lead in a very natural way to a system of three autonomous evolution equations that is similar to those obtained with ``expansion normalized'' variables for other known space--times (FLRW, Bianchi models, Kantowsky--Sachs, see \cite{EW}). 

Initial conditions (section VI) are selected from two ``primitive'' initial value functions (density, scalar 3--curvature) defined along an initial and regular hypersurface, $\T_i$ marked by constant comoving time. These functions determine the initial values for the phase space variables in such a way that the two spacelike constraints characterizing the 3+1 evolution equations for the LTB solutions are satisfied at the $\Ti$ and at subsequent $\T$. A third initial function determines the number of Regular Symmetry Centers (RSC) at the $\Ti$ (see Appendix A), and thus it determines the topological class of the $\T$, leading to hypersurfaces with ``open'' topology, (homeomorphic to $\mathbb{R}^3$, one RSC); ``closed'' topology (homeomorphic to $\mathbb{S}^3$, two RSC) ``wormhole'' (homeomorphic to $\mathbb{S}^2\times \mathbb{R}$ or to $\mathbb{S}^2\times \mathbb{S}^1$, zero RSC).

The resulting 3--dimensional phase space is discussed in section  VII, with its critical points, invariant subsets and particular solutions given in section VIII. This space is parametrized by three functions: a dimensionless shear scalar, $S$, a density average  parameter, $\Omav$, and a density contrast function $\Dm$. As long as initial conditions are selected so that unphysical shell crossing singularities do not arise~\cite{SG,HM,HL}, these variables are either bounded or can be trivially compactified for all configurations. These functions have a straightforward physical interpretation: the average $\Omav$ behaves exactly as the standard $\Omega$ in a FLRW dust spacetime: its initial values $\Omiav$ determine the dynamical evolution of each solution curve: re--collapse (if $\Omiav -1>0$) or perpetual expansion (if $\Omiav-1\leq 0$). Thus, $\Omav$ defines invariant subspaces $\Omav=0$,\, $\Omav=1$,\, $0<\Omav<1$ and $\Omav>1$. The density contrast function $\Dm$ provides a non--local characterization of the  type of inhomogeneity along the $\T$ (rest frames of fundamental observers):  a ``clump'' if $-1<\Dm\leq 0$  or a ``void'' if $\Dm\geq 0$, with $\Dm=0$ at the RSC. This variable also defines an invariant subset: $\Dm=-1$, corresponding to vacuum Schwarzschild--Kruskal solutions. Another invariant subspace is given by the line $S=\Dm=0$ marking the homogeneous and isotropic FLRW subcase, containing as subcases the Einstein--de Sitter and Minkowski--Milne universes.

We examine in detail in section IX the phase space evolution of several representative dust configurations: low density perpetually expanding (``hyperbolic'' dynamics) with $\Omiav-1\leq 0$, as well as a case of void formation scenario in which an initial density clump becomes a void (a numeric realization of configurations proposed in \cite{mustapha}). We also examine configurations with high density re--collapsing (``elliptic'') dynamics with open, closed and wormhole topologies ($\Omiav-1>0$); a ``structure formation'' scenario: solution curves near the RSC re--collapse into a black hole $\Omiav-1>0$ while ``external'' curves perpetually expand into a cosmic background $\Omiav-1\leq 0$. We also examine the vacuum subcase, which is the Schwarzschild--Kruskal spacetime given in comoving non--static coordinates made up by radial timelike geodesics.

All solution curves (except the RSC) for all configurations  start their evolution at a past attractor (source) at $\Omav=1,\,S=1/2,\,\Dm=-1$ that represents an initial big bang like singularity and is associated with a self similar regime. The evolution of each solution curve depends on the sign of $\Omiav-1$ and is fully contained in the invariant subsets associated with $\Omav$. Curves with $\Omiav-1=0$ (``parabolic'' evolution) remain in the plane $\Omav=1$ and terminate in a future attractor (sink) associated with the Einstein--de Sitter zero spacial curvature FLRW universe. Curves with $\Omiav-1< 0$ (``hyperbolic'' evolution) evolve towards a future attractor given by a line of sinks at $\Omav=S=0$, while curves with $\Omiav-1> 0$ (``elliptic'' re--collapsing evolution) evolve into a maximal expansion state $\Omav\to\infty$ and then terminate at the same critical point from which they started, though this point is now a sink or a future attractor. 

The RSC's of the configurations evolve separately from the rest of the curves along a line with $\Dm=S=0$, starting always from a source at $\Omav=1$ and going towards a sink at $\Omav=0$ or towards $\Omav\to\infty$, respectively, for expanding and re--collapsing configurations.  In all cases some of the curves pass near a saddle close to the RSC associated with homogeneity ($S=\Dm=0$), while in the structure formation scenario some curves also approach another saddle point that splits collapsing curves $\Omiav>1$ from expanding ones $\Omiav\leq 1$. The solution curves of the vacuum case are confined to the plane $\Dm=-1$, evolving towards $\Omav=0$ or $\Omav\to\infty$, depending on whether the radial geodesics are bound or not. 

We summarize and discuss the results from the article in section X. We also provide three appendices dealing with important issues: RSC's and geometric properties of hypersurfaces $\T$ (Appendix A), treatment of evolution equations given as PDE as a restricted dynamical system (Appendix B) and analytic solutions in terms of the variables used in the article (Appendix C).         

\section{LTB dust models in their original variables.}

The Lema\^\i tre--Tolman--Bondi (LTB) metric~\cite{kras,SG,HM,HMM} is the
spherically symmetric  line element
\begin{equation}ds^2=-c^2dt^2+\frac{Y'{}^2}{1-K}\,dr^2+Y^2(d\theta^2+\sin^2\theta
d\phi^2),\label{LTB1}\end{equation}
where $Y=Y(t,r)$,\,\ $Y'=\partial Y/\partial r $ and $K=K(r)$.
The momentum-energy tensor usually associated with (\ref{LTB1}) is that of a dust source:
\begin{equation} T^{ab}=\rho c^2 u^a u^b,\label{TabDust}\end{equation}
where $\rho=\rho(t,r)$ is the rest--mass density (in gm $\textrm{cm}^{-3} $).   

For the metric (\ref{LTB1}) and source (\ref{TabDust}) with a
comoving 4--velocity $u^a=\delta^a_0$,  Einstein's  field equations $G^a\,_b = \kappa
T^a\,_b$, with
$\kappa=8\pi G/c^4$, yield:
\ba \dot Y^2 &=& \frac{2M}{Y}-K,\label{Ysq_dust}\\
2M' &=& \kappa \rho c^2\, Y^2\,Y',\label{rho_dust}\ea
where $M=M(r)$ has units of length and $\dot Y=\partial Y/\partial x^0$.

The only nonzero
kinematic parameters are the expansion scalar $\Theta = u^a\,_{;a}$ and the shear 
tensor: $\sigma_{ab}=\nabla_{(a;b)} u-(\Theta/3) h_{ab}$, given by
\begin{equation}\Theta = \frac{2\dot Y}{Y}+\frac{\dot
Y'}{Y'},\label{Theta_}\end{equation}
\ba \sigma^a\,_{b} &=&
\textrm{{\bf diag}}\,[0,-2\Sigma,\Sigma,\Sigma],\nonumber\\
\Sigma &=& \frac{1}{3}\left[\frac{\dot Y}{Y}-\frac{\dot Y'}{Y'}\right].\label{Sigma_}
\ea
Other important quantities are the ``Electric'' Weyl tensor: $E_{ab}=C_{abcd}u^cu^d$  and the 3--dimensional Ricci tensor, $\RR$, of the hypersurfaces orthogonal to $u^a$ (marked by constant values of $x^0=ct$):
\ba E^a\,_{b} &=& \textrm{{\bf diag}}\,[0,-2\EE,\EE,\EE],\nonumber\\
\EE &=& \frac{M'}{3Y^2Y'}-\frac{M}{Y^3} = \frac{\kappa}{6}\,\rho c^2-\frac{M}{Y^3}.
\label{EWeyl}
\ea
\begin{equation}2(KY){}\,' =\RR \,Y^2Y'.\label{3ricci}\end{equation}
The usual treatment of LTB dust solutions consists in solving analytically the
evolution equation (\ref{Ysq_dust}). The solutions are usually classified in terms of the sign of $K$, which determines the type of evolution of dust layers: perpetually expanding/collapsing ``parabolic'' ($K=0$) and ``hyperbolic'' ($K<0$), and re-collapsing ``elliptic'' ($K>0$). These analytic solutions are given in Appendix C. However, instead of using these solutions, we will follow in this article a different approach based on defining new  variables more suitable for a qualitative and numerical
analysis. 

\section{Volume averages and scaling laws}

The form of $\rho$ and $\RR$ in terms of $2M'$ and $(KY){}\,'$ in (\ref{rho_dust})
and (\ref{3ricci}) suggests considering the following volume averages along an
arbitrary hypersurface $\T_i$ orthogonal to $u^a$ and marked by constant
$x_i^0=ct_i$
\begin{equation}\langle A_i\rangle \equiv \frac{\int{A_i\,
d\VV_i}}{\int{d\VV_i}},\qquad
d\VV_i=Y_i^2Y_i'\sin\theta\,dr\,d\theta\,d\phi,\label{vol_ave}\end{equation}
where the subindex ${}_i$ denotes evaluation at $t=t_i$ and we have assumed that $\T_i$
is fully regular in the integration range. Applying (\ref{vol_ave}) to (\ref{rho_dust})
and (\ref{3ricci}) we get
\ba \frac{\kappa c^2}{3}\rhoav = \frac{\kappa c^2}{3}\,\frac{\int{\rho\,Y^2\,Y'
dr}}{\int{Y^2\,Y'dr}}=\frac{2M}{Y^3},\label{rhoave}\\
\frac{1}{6}\RRav=\frac{1}{6}\,\frac{\int{\RR\,Y^2\,Y'
dr}}{\int{Y^2\,Y'dr}}=\frac{K}{Y^2},\label{3Rave}\ea
where we have dropped the subindex ${}_i$ and have taken the lower bound of
the integration range of $r$ to be determined by a suitable boundary condition on $\rho_i$ and $\RRi$, for example a RSC. See Appendix A and section IX--E.

The averages $\rhoav$ and $\RRav$ are non--local quantities depending on the form of
$\rho$ and $\RR$ along the integration range. We define the following ``contrast
functions'' comparing these quantities with their local counterparts:
\ba \Dm &\equiv& \frac{\rho-\rhoav}{\rhoav}, \quad\Rightarrow\quad \rho=\rhoav
[1+\Dm],\label{Dm}\\
\Dk &\equiv& \frac{\RR-\RRav}{\RRav}, \quad\Rightarrow\quad \RR=\RRav
[1+\Dk],\nonumber\\\label{Dk}\ea
The interpretation of these these contrast functions~\cite{SG} follows by using the definitions
(\ref{rho_dust}), (\ref{rhoave}) and (\ref{Dm}) and integrating by parts along the $\T$, 
leading to
\begin{equation} \Dm = \frac{\kappa c^2}{6M}\int{\rho\,'\,Y^3\,dr},\label{interp1}
\end{equation}
where this integral is evaluated from a RSC. Since $M\geq 0$ and $Y\geq 0$ and assuming $\rho\geq 0$, the only quantity that can change sign is $\rho'$, therefore, looking at the radial variation of a density profile from the symmetry center along an arbitrary $\T$ we have
\begin{equation}
\Dm   \left\{ \begin{array}{l}
  \le 0\,  \Leftrightarrow \, \rho\,'  \le 0,\,\,\, \rho    \le \rhoav ,\quad
{\hbox{density clump}} \\ 
  \ge 0\,  \Leftrightarrow \, \rho\,'  \ge 0,\,\,\, \rho    \ge \rhoav ,\quad
{\hbox{density void}} \\ 
 \end{array} \right.\label{Dm_CV}
\end{equation}

Likewise, for the contrast function $\Dk$, using the definitions (\ref {3ricci}), 
(\ref{3Rave}) and (\ref{Dk}) and integrating by parts along the $\T$ we obtain an
analogous expression to (\ref{interp1}):
\begin{equation} \Dk = \frac{1}{6K}\int{\RR'\,Y^3\,dr},\label{interp2}
\end{equation}
However, while a negative $\rho$ is not physically interesting, there is no physical objection
for $\RR$ being positive or negative, or changing sign in the allowed range of $r$. If
$\RR> 0$ for all the allowed range of $r$, we have the same results as with $\Dm$ in
(\ref{Dm_CV}), but if $\RR< 0$, then curvature clumps or voids are defined by the opposite signs~\cite{SG}. 

We define now the following ``scale factors'' related to the metric functions:
\ba \ell &\equiv& \frac{Y}{Y_i},\label{ell}\\
\Gamma &\equiv& \frac{Y'/Y}{Y_i'/Y_i} = 1+\frac{\ell\,'/\ell}{Y_i'/Y_i},
\label{Gamma}\ea
relating $Y$ and $Y'$ evaluated at an arbitrary and a fiducial (or ``initial'') 
hypersurface $\T_i$. Since (\ref{rhoave}) and (\ref{3Rave}) are valid at any $\T$, we
obtain the scaling laws
\ba \rhoav &=& \frac{\rhoiav}{\ell^3},\label{rhoaveSL}\\
\RRav &=& \frac{\RRiav}{\ell^2}.\label{RRaveSL}\ea
while from (\ref{rho_dust}), (\ref{3ricci}) and (\ref{ell})--(\ref{Gamma}) we
get
\ba \rho &=& \frac{\rho_i}{\ell^3\,\Gamma},\label{rho1}\\
\RR &=& \frac{1}{\ell^2\,\Gamma}\,\left[\RRi
+\frac{1}{3}\RRiav\,(1-\Gamma)\right],\label{RR1}\ea
Comparing (\ref{Dm})--(\ref{Dk}) with (\ref{rhoaveSL})--(\ref{RRaveSL}) and
(\ref{rho1})--(\ref{RR1}) we get the scaling laws for $\Dm$ and $\Dk$
\ba 1+\Dm &=& \frac{1+\Dim}{\Gamma},\label{DmSL}\\
\frac{2}{3}+\Dk &=& \frac{2/3+\Dik}{\Gamma},\label{DkSL}\ea

The quantities $\Theta$, $\Sigma$ and $\EE$ given in (\ref{Theta_}), (\ref{Sigma_})
and (\ref{EWeyl}) take the forms
\ba \Theta &=& \frac{3\dot \ell}{\ell}+\frac{\dot
\Gamma}{\Gamma},\label{Theta}\\
\Sigma &=& -\frac{\dot \Gamma}{3\Gamma},\label{Sigma}\\
\EE &=& \frac{\kappa c^2}{6}\left[\ \rho-\rhoav \ \right]=\frac{\kappa
c^2}{6}\rhoav\,\Dm,\label{E}\ea
the evolution equation (\ref{Ysq_dust}) becomes 
\begin{equation} \frac{\dot \ell^2}{\ell^2} \ = \ \frac{\kappa c^2}{3}
\rhoav -
\frac{1}{6}\RRav,\label{eveq}\end{equation}
while the LTB metric (\ref{LTB1}) takes the Friedmanian form
\begin{widetext}
\begin{equation} ds^2 \ = \
-c^2dt^2+\ell^2\left[\frac{\Gamma^2\,Y_i'{}^2\,dr^2}{1-\frac{1}{6}
\RRiav Y_i^2}+Y_i^2\,\left(d\theta^2+\sin^2\theta\,d\phi^2\right)\right],\label{LTB2}
\end{equation}
\end{widetext}

It is also helpful to express the spacial ({\it i.e. radial}) gradients of $\rhoav$ and
$\RRav$ in terms of the contrast functions and $\Gamma$.  With the help of (\ref{rho_dust}),
(\ref{3ricci}), (\ref{rhoave}), (\ref{3Rave}), (\ref{Dm}) and (\ref{Dk}) we obtain
the following useful relations:
\ba \frac{\rhoav'}{\rhoav} &=& \frac{3Y'}{Y}\,\Dm =
\frac{3Y_i'}{Y_i}\,\Dm\,\Gamma,\label{grad1}
\\
\frac{\RRav'}{\RRav} &=& \frac{3Y'}{Y}\,\Dk = 
\frac{3Y'}{Y}\,\Dk\,\Gamma,\label{grad2}\ea
Notice that these equations are strictly valid along all hypersurfaces $\T$.

\section{3+1 decomposition.}

From a covariant 3+1 decomposition of spacetime based on the 4--velocity field $u^a$ presented in~\cite{EW,EVE}, the field and energy balance equations for a dust model characterized by (\ref{LTB1}) and
(\ref{TabDust}) are equivalent to the following set of first order evolution 
equations:
\bse\label{system1}\ba \frac{\dot
\Theta}{3} &=& -\left(\frac{\Theta}{3}\right)^2-\frac{1}{3}\sigma_{ab}\,
\sigma^{ab}-\frac{\kappa c^2}{6}\rho=0,\label{ev_Theta}
\\
\dot\sigma^{\langle ab \rangle} &=& -\frac{2}{3}\Theta\,\sigma^{ab}-\sigma^{\langle
a}{}_{c}\,\sigma^{b\rangle c}-E^{ab}=0,\label{ev_sigma}
\\
\dot  E^{\langle ab \rangle} &=& -\frac{\kappa
c^2}{2}\rho\,\sigma^{ab}-\Theta\,E^{ab}+3\sigma^{\langle a}{}_c\,E^{b\rangle c},
\label{ev_E}
\\
\dot\rho &=& -\rho\Theta,\label{ev_rho}\ea\ese
together with the constraints:
\bse\ba \tilde \nabla_b\,\sigma^b\,_{a}
-\frac{2}{3}\tilde\nabla_a\,\Theta=0,\label{constr}\\
\tilde \nabla_b\,E^b\,_{a}
-\frac{\kappa c^2}{3}\tilde\nabla_a\,\rho=0, \label{constrE}\ea\ese
where $\dot A^{ab}\equiv u^c\nabla_c A^{ab}$ is the convective derivative along  the
4--velocity, $\tilde \nabla_a A^{bc} \equiv h_a{}^d\nabla_d A^{bc}$ is the spacelike
gradient, tangent to the $\T$ hypersurfaces and orthogonal to $u^a$, and
$A^{\langle ab\rangle}\equiv \,A^{(ab)}-(1/3)A^c{}_c h^{ab}$ is the spacelike,
symmetric, trace--free part of a tensor 
$A^{ab}$.

Equations (\ref{ev_Theta})--(\ref{ev_rho}) are 4 tensorial equations for 4 tensorial quantities
$\Theta,\,\rho,\,\sigma^a{}_b,\,E^a{}_b$, which together with the constraints
(\ref{constr})--(\ref{constrE}) form a  completely determined system of partial differential
equations (equivalent to Einstein's field equations). 
Bearing in mind that $u^a=\delta^a{}_0$ and considering the forms of the trace--free tensors $\sigma^a{}_{b}$ and $E^a{}_{b}$ in (\ref{Sigma_}) and (\ref{EWeyl}), the system (\ref{system1}) reduces to the following set of scalar
equations:
\bse\label{system2}\ba \frac{\dot
\Theta}{3} &=& -\left(\frac{\Theta}{3}\right)^2-2\Sigma^2-\frac{\kappa
c^2}{6}\rho=0,\label{ev_Theta2},
\\
\dot\Sigma &=& -\frac{2}{3}\Theta\,\Sigma+\Sigma^2-\EE,\label{ev_sigma2}
\\
\dot\EE &=& -\EE-\,\Sigma\left[\frac{\kappa}{2}\,\rho c^2+3\EE\right],\label{ev_E2}
\\
\dot\rho &=& -\rho\Theta,\label{ev_rho2}\ea\ese
and (\ref{constr})--(\ref{constrE}) become
\bse\label{constr20}\ba\Sigma\,'+\frac{\Theta'}{3}+3\,\Sigma\,\frac{Y'}{Y}=0,\label{constr2}
\\
\EE\,'+\frac{\kappa\rho\,'\,c^2}{6}+3\,\EE\,\frac{Y'}{Y}=0.\label{constr2E}\ea\ese

It is straightforward to express the system (\ref{system2}) and the constraints
(\ref{constr20}) in therms of the new variables introduced in the previous
sections. Considering (\ref{Dm}), (\ref{DmSL}) and (\ref{Theta})--(\ref{E}) and and defining
\begin{equation}H\equiv \frac{\dot \ell}{\ell}=\frac{\Theta}{3}+\Sigma,
\label{H_def}\end{equation}
the system (\ref{system2}) becomes the following set of equivalent
evolution equations:
\bse\label{system3}\ba \dot H &=& -H^2-\frac{\kappa}{6}\rhoav,\label{ev_H3}
\\
\dot \Sigma &=& 3\Sigma^2-2H\Sigma+\frac{\kappa}{6}\rhoav\,\Dm,\label{ev_Sigma3}
\\
\rhoav \dot{} &=& -3\,\rhoav\,H,\label{ev_rhoav3}
\\
\dot\Dm &=& 3[1+\Dm]\,\Sigma,\label{ev_Dm3}\ea\ese
while the constraints (\ref{constr2E}) and (\ref{constr2}) become respectively
\bse\ba \rhoav \,' - 3\rhoav\,\Dm\,\Gamma\frac{Y_i'}{Y_i} = 0, \label{constr3E}\\
H' - 3\,\Sigma\,\frac{Y'}{Y}=H' - 3\,\Sigma\,\Gamma\,\frac{Y_i'}{Y_i}
=0.\label{constr3}\ea\ese 

The constraint (\ref{constrE}) leads to (\ref{constr3E}) which is identical  to
(\ref{grad1}), so this constraint is automatically satisfied at all $t$ (all $\T$) by a
system of evolution equations based on the variables $\rhoav$ and $\Dm$. 

Bearing in mind (\ref{ell}), (\ref{Gamma}) and (\ref{Sigma}),  the constraint
(\ref{constr}) in its form (\ref{constr3}) can be written as the integrability  condition
for $\ell$:
\begin{equation}\Gamma\left(\frac{\dot\ell}{\ell}\right)'
-\dot\Gamma\left(\frac{Y_i'}{Y_i}+\frac{\ell\,'}{\ell}\right)=\Gamma\,
\left[\left(\frac{\dot\ell}{\ell}\right)'-\left(\frac{\ell\,'}{\ell}\right)
\dot{}\right]=0.
\label{int_cond1}\end{equation}
Further, since (\ref{int_cond1}) holds and (\ref{H_def}) implies
\begin{equation}H=\frac{\dot\ell}{\ell}=\left[\frac{\kappa
c^2}{3}\,\rhoav-\frac{1}{6}\RRav\right]^{1/2},\label{H_def3}\end{equation}
the constraint (\ref{constr3}) implies (with the help of (\ref{DmSL}), (\ref{DkSL}), (\ref{grad1}), 
and (\ref{grad2})):
\begin{equation}\Sigma = -\frac{1}{2}\,\frac{\frac{1}{3}\kappa c^2\rhoav
\Dm-\frac{1}{6}\RRav\Dk }{\left[\frac{1}{3}\kappa
c^2\rhoav-\frac{1}{6}\RRav\right]^{1/2}}.\label{Sig3}\end{equation}
If the system (\ref{system3}) is self--consistent,  then (\ref{Sig3}) must
hold for every $\T$ and so it must comply with (\ref{ev_Sigma3}). Inserting (\ref{Sig3})
into (\ref{ev_Sigma3}) and eliminating time derivatives of $\rhoav,\,\Dm,\,H,\,\RRav$ and
$\Dk$ with the help of (\ref{ev_H3}), (\ref{ev_rhoav3}),  (\ref{ev_Dm3}) and the scaling
laws (\ref{DmSL})--(\ref{DkSL}), it is straightforward to see that (\ref{ev_Sigma3})
is identically satisfied at every $\T$.

\section{Dimensionless variables: a dynamical system} 

In order to transform system (\ref{ev_H3})--(\ref{ev_Dm3}) into a dynamical system, we need to recast $H,\,\Sigma$ and $\rhoav$ in terms of dimensionless ``expansion normalized'' variables.~\cite{EW}. However, instead of using the expansion parameter $\Theta$ in (\ref{Theta}), it turns out to be easier to use $H$ defined in (\ref{H_def}). This leads to:
\ba \HH &\equiv& \frac{H}{H_0},\label{CH_def}
\\
\Omav &\equiv& \frac{\kappa c^2\,\rhoav}{3 H^2},\label{Omega_def}
\\
S &\equiv& \frac{\Sigma}{H},\label{S_def}
\ea
where $H_0$ is a characteristic length scale. Notice that we can also define a dimensionless $\Omega$ parameter for
the local density, which would be related to that of (\ref{Omega_def}) by:
\begin{equation}\Omega = \frac{\kappa c^2\,\rho}{3 H^2} =
\Omav [1+\Dm].\label{Omega_def2}\end{equation}
We also need to make the ``dot'' derivative $\partial/\partial ct $ a
dimensionless operator, so we define:
\begin{equation}\frac{\partial}{\partial \tau} \equiv \frac{1}{H_0}
\frac{\partial}{c\partial t}.\label{tau_def}\end{equation}

Inserting (\ref{CH_def})--(\ref{tau_def}) into (\ref{ev_H3})--(\ref{ev_Dm3})  we obtain the
following dimensionless system:
\bse\label{system4}\ba  \HH_{,\tau} &=&
-\left[\,1+\textstyle{\frac{1}{2}}\Omav\right]\,\HH^2,\label{ev_CH4}
\\
S_{,\tau} &=&
\HH\left[\,S(3S-1)+\textstyle{\frac{1}{2}}\left(\Dm+S\right)\,
\langle\Omega\rangle\right],\label{ev_S4}
\\
\Omav_{,\tau} &=&
\HH\,\Omav\,\left[\,\Omav-1\,\right],\label{ev_Omega4}
\\
\Dm_{,\tau} &=& 3\HH\, S\,\left[\,1+\Dm\,\right].\label{ev_Dm4}\ea\ese
where $_{,\tau}=\partial/\partial \tau$.

The form of these equations suggests that further simplification follows if the three equations (\ref{ev_S4})--(\ref{ev_Dm4}) can be decoupled from (\ref{ev_CH4}). Defining the following coordinate transformation:
\begin{equation} \tau = \tau(\xi, \bar r),\qquad r = \bar r\end{equation}
so that for any function $A=A(\tau,r)$ we have $A(\tau,r)=A(\xi(\tau,\bar r),\bar r)$ and
all partial time drivatives in (\ref{system4}) can be expressed as 
\begin{equation}\left[\frac{\partial A}{\partial \tau}\right]_r = \frac{\partial
A}{\partial \xi}\,\left[\frac{\partial \xi}{\partial \tau}\right]_r = \HH\, \frac{\partial
A}{\partial \xi},\label{chvar1}\end{equation}
where $\xi$ is selected so that:
\begin{equation}\left[\frac{\partial \xi}{\partial \tau}\right]_r =
\HH,\quad\Rightarrow\quad \xi = \ln\ell,\label{chvar2}\end{equation}
The introduction of the new variable $\xi$ defined by (\ref{chvar1}) and (\ref{chvar2}) removes the dependence on $\HH$ of the evolution equations (\ref{ev_S4}), (\ref{ev_Omega4})
and (\ref{ev_Dm4}): 
\bse\label{system5}\ba \frac{\partial S}{\partial \xi} &=&
S\left(3\,S-1\right)+\frac{1}{2}\left(\Dm+S\right)\,\Omav,\label{ev_S5}
\\
\frac{\partial \Dm}{\partial \xi} &=& 3\,S\,\left[\,1+ \Dm \right],\label{ev_Dm5}
\\
\frac{\partial \Omav}{\partial \xi} &=&
\Omav\,\left[\Omav-1\right],\label{ev_Omega5}\ea\ese
while (\ref{ev_CH4}) becomes:
\begin{equation} \frac{\partial}{\partial\xi}(\ln \HH)  = -\left[1+\frac{1}{2}\Omav
\right],\label{ev_CH5}\end{equation} 

The system (\ref{system5}) is formally a dynamical system constructed with ``expansion normalized'' variables~\cite{EW}. However, it is still a system of PDE's, even if it only contains derivatives with respect to $\xi$ and $r$ is basically a parameter. As we show in Appendix B, this system of PDE's equivalent authonomous ODE system with restricted initial conditions (restricted in order to fulfill the spacelike constraints). 

\section{Initial conditions}

Bearing in mind (\ref{rho_dust}), (\ref{3Rave}), (\ref{Dm})--(\ref{Dk}) and
(\ref{H_def3})--(\ref{Sig3}),  intial conditions for the system
(\ref{system5}) can be constructed through the following steps:

\begin{enumerate}

\item Choose the topological class of the initial hypersurface $\T_i$ in terms of the number of RSC (see Appendix A). A convenient choice is:
\begin{equation}Y_i=H_0^{-1} f(r),\label{initfuns}
\end{equation}
where $f(r)$ is a (at least a $C^2$) function whose zeros correspond to RSC's and $H_0$ is the same characteristic length scale as in (\ref{CH_def}) and (\ref{tau_def}). Strictly speaking, the choice of $f$ is a choice of radial coordinate (see Appendix A).
\item Construct dimensionless quantities out of $\rho_i$ and $\RRi$:
\begin{equation}m_i \equiv \frac{\kappa c^2\,\rho_i}{3H_0^2},\qquad
k_i=\frac{\RRi}{6H_0^2},\label{no_dim}\end{equation}
\item Obtain the orbit volume averages and contrast functions
\bse\ba \langle m_i\rangle &=& \frac{\kappa c^2}{3H_0^2}\,\rhoiav,\label{no_dim_rhoiav}
\\
\langle k_i\rangle &=& \frac{1}{6 H_0^2}\,\RRiav,\label{no_dim_RRiav}
\\
 \Dim &=& \frac{m_i}{\langle m_i\rangle}-1,\label{Dm5}
\\
\Dik &=& \frac{k_i}{\langle k_i\rangle}-1,\label{Dk5}\ea\ese
\item The initial forms of the remaining variables are then:
\bse\label{infuncs}\ba 
\HH_i &=& \left[\,\langle m_i\rangle-\langle k_i\rangle\right]^{1/2},
\label{HHi}
\\
\Omiav &=&
\frac{\langle m_i\rangle}{\langle m_i\rangle-\langle k_i\rangle},
\label{Omegaavi}
\\
S_i &=&
-\frac{1}{2}\,\frac{\langle m_i\rangle\,\Dim-\langle k_i\rangle\,\Dik}
{\langle m_i\rangle-\langle k_i\rangle},\label{Si}
\ea\ese

\end{enumerate}

Just as we proceeded with
(\ref{system3}), the fulfillment of constraints (\ref{constr3E}) and (\ref{constr3}) imply that
the functional forms of $\HH$,\,$\Omav$ and $S$ are the generalization of
(\ref{infuncs}) for all times, that is:
\bse\label{funcs1}\ba 
\HH &=& \left[\,\langle m\rangle-\langle k\rangle\right]^{1/2},
\label{HH_}
\\
\Omav &=&
\frac{\langle m\rangle}{\langle m\rangle-\langle k\rangle},
\label{Omegaav_}
\\
S &=&
-\frac{1}{2}\,\frac{\langle m\rangle\,\Dm-\langle k\rangle\,\Dk}
{\langle m\rangle-\langle k\rangle},\label{S_}
\ea\ese
where
\bse\label{funcs2}\ba\mav &=& \frac{\kappa c^2\rhoav}{3
H_0^2} = \frac{\miav}{\ell^3},\\
 \langle k\rangle &=& \frac{\RRav}{6H_0^2}  =
\frac{\kiav}{\ell^2}.\ea\ese
Inserting (\ref{funcs1}) and (\ref{funcs2}) into (\ref{system5}) and eliminating derivatives of $\mav$ and $\kav$ by means of 
(\ref{DkSL}), (\ref{Sigma}), (\ref{ev_H3}), (\ref{ev_rhoav3}) and  (\ref{ev_Dm3}), we
can see that (\ref{funcs1}) and (\ref{funcs2}) fully solve (\ref{system4}). Thus
(\ref{system4}) propagates the initial data given by (\ref{initfuns})--(\ref{Si}) with the constraints (\ref{constr3E}) and (\ref{constr3}) satisfied for all $\T$.

The standard approach to LTB dust solutions is based on solving the evolution equation (\ref{Ysq_dust}). Therefore, it is illustrative to re--write such equation in terms of the initial value variables that we have defined here:
\begin{equation}\ell^2_{,\tau}=\HH_i^2\left[\frac{\Omiav}{\ell}-\left(\Omiav-1\right)\right].\label{ellsq_dust}\end{equation}  
This form of (\ref{Ysq_dust}) shows how the sign of $\Omiav-1$ determines the type of evolution of dust layers, just as the sign of $K$ does it in (\ref{Ysq_dust}), leading to ``parabolic'' ($\Omiav-1=0$), hyperbolic'' ($\Omiav-1<0$) and ``elliptic'' ($\Omiav-1>0$) evolution like setting $K=0,\,K<0,\, K>0$ in (\ref{Ysq_dust}). See \cite{HM,HMM}. Analytic solutions of (\ref{ellsq_dust}) are given in Appendix C.

In the numerical examination of dust configurations we will need to solve numerically system (\ref{system3}),  besides system (\ref{system5}). In terms of the variables and initial conditions (\ref{initfuns})--(\ref{funcs2}) the system (\ref{system3}) takes the dimensionless form:
\bse\label{sys3}\ba \HH_{,\tau} &=& -\HH^2-\frac{\mav}{2},
\\ 
s_{,\tau} &=& 3\,s^2-2\HH\,s+\frac{\mav}{2}\,\Dm,
\\
\mav_{,\tau} &=& -3\mav\,\HH,
\\
\Dm_{,\tau} &=& 3(1+\Dm)\,s,
\ea\ese
where $\tau$,\, $\HH$ and $\mav$ have been defined in (\ref{CH_def}), (\ref{tau_def}),\, (\ref{funcs2}) and $s=\Sigma/H_0$. This system will be specially useful for looking at the collapsing stage of re--collapsing configurations, a feature that cannot be studied by (\ref{system5}) because maximal expansion takes place as $\HH\to 0$,  and so $\Omav$ diverges and the solution curves cannot be extended further to study their collapsing stage. Also, (\ref{sys3}) yields directly quantities like $\mav$ and $\HH$, while the metric functions $\ell$ and $\Gamma$ in (\ref{LTB2}) follow from the variables in (\ref{sys3}) as:
\begin{equation}\ell = \left[\frac{\miav}{\mav}\right]^{1/3},\quad \Gamma=\frac{1+\Dim}{1+\Dm},\end{equation}
and the so--called ``curvature radius'' $Y=\sqrt{g_{\theta\theta}}$ can be easily found as $Y=Y_i\,\ell$. 
    
\section{Some qualitative and analytic results.} 

The solution curves $[S(\xi,r),\Dm(\xi,r),\Omav(\xi,r)]$ of (\ref{system5}) evolve in a phase space which is a region of $\mathbb{R}^3$ parametrized by the coordinates $[S,\Dm,\Omav]$. The evolution variable is $\xi$, while each curve represents a fundamental comoving observer labelled by a constant value $r$. It is highly desirable that the phase space coordinates remain bounded as these curves evolve along their maximal range of $\xi$ for all $r$. We discuss in this section several useful analytic results and the relation between the phase space variables and their initial values. 

Considering that  $\xi=\ln\ell$ and 
\bse\ba \Omiav-1  &=& \frac{\kiav}{\miav-\kiav},\\
\Omav-1  &=& \frac{\kav}{\mav-\kav} = \frac{\kiav\,\textrm{e}^\xi}{\miav-\kiav\,\textrm{e}^\xi}, \ea\ese
the functions $\HH,\,\Omav$ and $S$ in (\ref{funcs1})--(\ref{funcs2}) can be given as 
\bse\label{funcs_Om}\ba \Omav &=& \frac{\Omiav}{\Omiav-[\Omiav-1]
\textrm{e}^\xi},
\label{Ome_ana}\\
\HH^2 &=& \HH_i^2\left[\Omiav
\,\textrm{e}^{-3\xi}-(\Omiav-1)\,\textrm{e}^{-2\xi}\right],\label{H_ana}
\\S &=&
-\frac{\Omiav\Dm-[\Omiav-1]\,\Dk\,\textrm{e}\,^\xi}
{2\left[\Omiav-[\Omiav-1]\,\textrm{e}\,^\xi\right]}.\label{S_ana}
\ea\ese 
where we are emphasizing that these  are now functions of $(\xi,r)$. Notice that even if, in general, the surfaces of constant $\xi$ do not
coincide with the $\T$ (surfaces of constant $t$ or $\tau$),  the initial hypersurface
$\xi=0$ does coincide with the initial $\Ti$. 

\subsection{Constraints on $\Omav$}

From(\ref{funcs1}), we have the following important constraint:
\begin{equation} \Omav - 1 \ = \ \frac{\langle k \rangle }{\HH^2}\ = \ \frac{\langle k_i \rangle}{\ell^2\,\HH^2}\ = \ \frac{\Omiav -1}{\ell^2\,\HH^2}.\label{Hconstr} \end{equation}
Since $\Omav \geq 0$, this constraint (together with (\ref{Ome_ana})) has important qualitative consequences: it defines invariant subsets associated with $\Omav$. If initial conditions for a range of $r$ are selected so that $\Omiav =0,\,0<\Omiav <1,\, \Omiav=1$, or $\Omiav > 1$, then all solution curves for such range respectively comply with $\Omav =0,\,0<\Omav <1,\, \Omav=1$, or $\Omav > 1$ for all their maximal extensibility range of of the evolution parameter $\xi$.

Notice from (\ref{funcs_Om}) that,
irrespectively of the sign of $\Omiavr-1$, we have $\Omav\to 1$ as $\xi\to\ -\infty$ (or, equivlently, as $\ell\to 0$). Thus, irrespectively of the type of dynamics in(\ref{H_def3}) given by the sign of $\Omiav -1 $, all solution curves  of (\ref{system5}) will approach the plane $\Omav = 1$ near the initial big bang singularity (in the asymptotic range: $\xi \to -\infty$ or $\ell\to 0$).

Since $\Omav \ge 0$, (\ref{Hconstr}) implies that all solution curves with initial conditions $0\leq \Omiav \leq 1$ will be constrained to the invariant subset given by the region $0 \leq \Omav \leq 1$ for all values of $\xi$, while (\ref{funcs_Om}) implies that the such curves will evolve from the plane $\Omav= 1$ in $\xi\to-\infty$ ($\ell\to 0$) towards the plane $\Omav= 0$ in the asymptotic range $\xi\to\infty$ (or $\ell\to \infty$).  

Equations (\ref{funcs_Om}) place no restriction in the range of $\xi$ of solution curves when $\Omiav\leq 1$, but curves with initial conditions given by $ \Omiav > 1$ the evolution parameter $\xi$ will have a bounded range of maximal extendibility given by
\begin{equation}\langle
\Omega\rangle\to\infty,\quad\textrm{as}\quad
\xi\to\ln\frac{\Omiav}{\Omiav
-1}.\label{maxexp}\end{equation}
Such solution curves start from the plane $\Omav = 1$ as $\xi\to -\infty$ ($\ell\to 0$) and diverge for the limiting value of $\xi=\ln\ell$ given by (\ref{maxexp}). However, this is the same finite  value of $\xi$ that makes $\HH$ in (\ref{H_ana}) vanish, so the blowing up $\Omav\to\infty$ of all solution curves in this case simply marks the values of $\ell$ associated with  the ``maximal'' expansion of dust layers (maximal value of $\ell$) in the ``turn around'' before the collapsing stage.  Since $\Omav$ blows up the collapsing stage cannot be described by (\ref{system5}) (though it can be described by (\ref{sys3})).

\subsection{Constraints on $S$ and $\Dm$}

In order to appreciate the relation between the coordinates of the phase space, $[S,\Dm, \Omav]$, it is very useful to express $S$ given by (\ref{S_}) as:
\begin{equation} S = -\frac{1}{2}\left\{\Omav\,\Dm-\left[\Omav-1\right]\,\Dk \right\},\label{S_constr}\end{equation}
where, from the scaling laws (\ref{DmSL}) and (\ref{DkSL}), we have 
\bse\label{DmkSL}\ba\Dm &=& \frac{1+\Dim}{\Gamma}-1,\\
\Dk &=& \frac{2/3+\Dik}{\Gamma}-\frac{2}{3}.\ea\ese
Thus, for finite $\Dim$ and $\Dik$, it is evident that a sufficient condition for $\Dm$ and $\Dk$ to remain bounded for all $\xi$ is that initial conditions are selected so that a shell crossing singularity does not emerge (see Appendix A and references~\cite{SG,HM,HMM,HL}). That is, we must have for all solution curves marked by $(\xi,r)$ 
\begin{equation}\Gamma(\xi,r) > 0,\qquad \textrm{no shell crossing singularity,}\label{nxcr}\end{equation}
where $\Gamma$ has been defined in (\ref{Gamma}). Notice, from (\ref{S_ana}), (\ref{S_constr}) and (\ref{DmkSL}), that even if (\ref{nxcr}) holds, $S$ (like $\Omav$) diverges as $\HH\to 0$ (maximal expansion) for curves with $\Omiav>1$, though it remains bounded if $\Omiav\leq 1$. In order to overcome the blowing up of $S$ and $\Omav$ in the cases when $\Omiav>1$ we have used the system (\ref{sys3}) to examine the collapsing stage of the solution curves. 

It is important to remark that for initial conditions for which both (\ref{nxcr}) and $0<\Omiav\leq 1$ hold, the three coordinates of phase space, $[S,\Dm,\Omav]$, of all solution curves remain bounded and restricted to a finite region of $\mathbb{R}^3$. 

Initial conditions that guarantee the fulfillment of (\ref{nxcr}) were derived in terms of original variables in reference~\cite{HL} (see also~\cite{HM,HMM}), in terms of initial value variables discussed here in~\cite{SG} (see Appendix A).

\section{Critical points and particular cases}

In the general case  in which none of the variables $[S,\,\Dm, \Omav]$ is restricted to take any special constant value associated with particular cases, the  coordinates and nature of critical points of (\ref{system5}) are:

\bi 

\item $\Omav =1 $:
\bse\label{crps1}\ba 
&{\bf{C_1}}&\quad S = 1/2,\quad \Dm =-1,\quad \hbox{source},\label{cr11}\\ 
&{\bf{C_2}}&\quad S = -1/3,\quad \Dm =-1,\quad \hbox{saddle},\label{cr12}\\
&{\bf{C_3}}& \quad S = 0,\qquad \Dm =0, \ \quad\hbox{saddle},\label{cr13}\ea\ese

\item $\Omav =0 $:
\bse\label{crps0}\ba  
&{\bf{C_4}}& \quad S = 1/3,\quad \Dm =-1,\qquad \hbox{saddle},\label{cr01}\\
&{\bf{C_5}}& \quad S = 0,\quad \Dm \ \hbox{arbitrary},\qquad \hbox{sink},\label{cr02}\\
&{\bf{C_6}}&\quad S=0,\quad\Dm=0,\qquad \qquad\hbox{sink},\label{cr03}
\ea\ese
\ei

\begin{figure}[htbp]
\begin{center}
\includegraphics[width=2.5in]{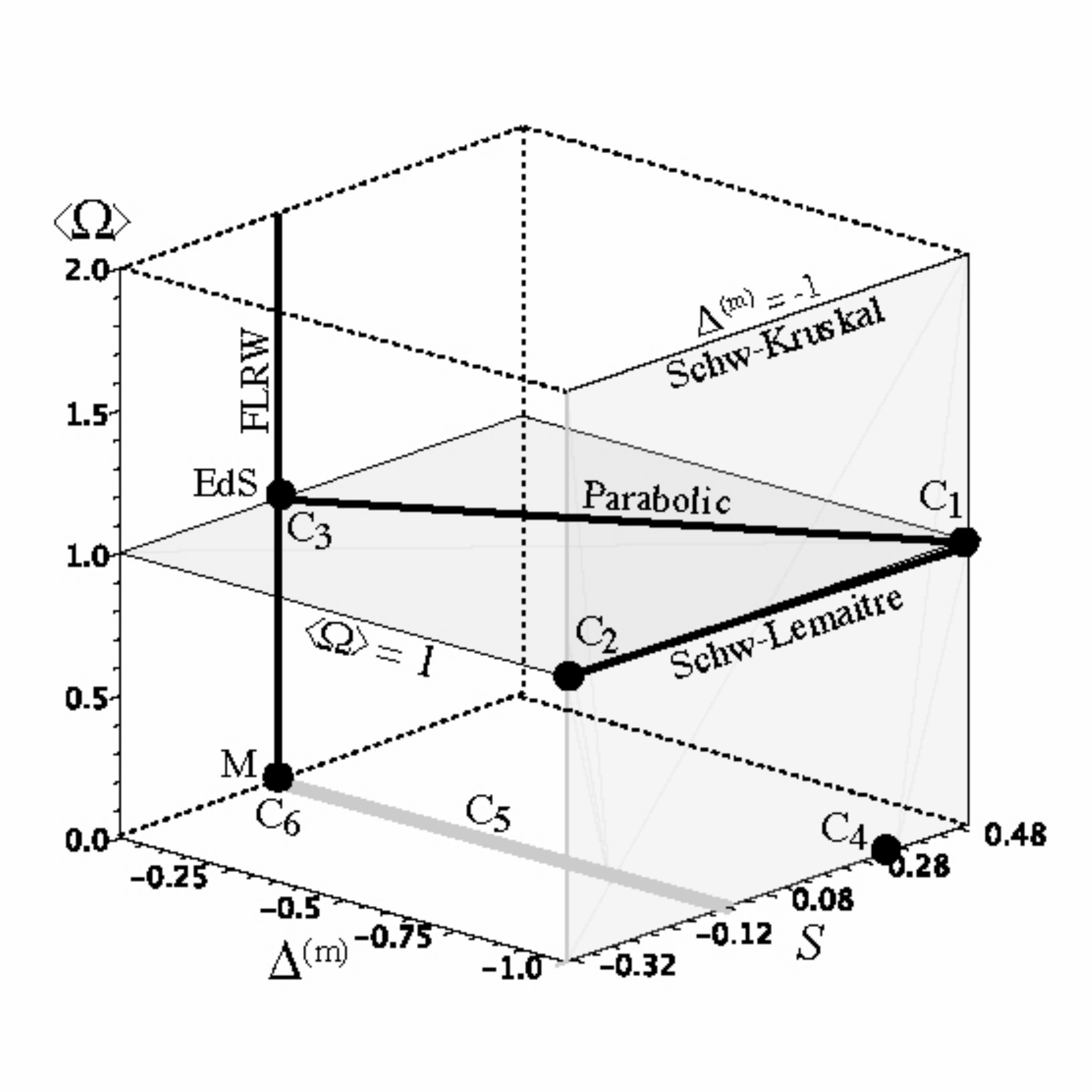}
\caption{{\bf Phase space} The figure shows all critical points and the invariant subspaces given by the surfaces $\Omav=1$ and $\Dm=-1$ (Schwarzschild vacuum case). Particular cases, such as FLRW, parabolic evolution and the vacuum case in Lema\^\i tre coordinates, are also invariant subspaces shown as lines going from a source to a sink. The FLRW sub--case contains the Einstein--de Sitter (EdS) universe $\Omav=1$ (the critical point ${\bf C_3}$) and the Minkowski--Milne space--time (M) $\Omav=0$ (the critical point ${\bf C_6}$).  }
\label{phase_space}
\end{center}
\end{figure}

The nature of some of the critical points of (\ref{system5})  changes when $[S,\Dm,\Omav]$ take certain specific particular values that correspond to various space--times that are particular cases of dust LTB solutions. We examine these particular cases and their critical points in the remaining of this section. The phase space and the location of critical points and location of all particular cases is depicted in figure \ref{phase_space},

\subsection{Parabolic (or ``marginally bound'') evolution}

For initial conditions $\Omiav =1$ equations (\ref{funcs_Om}) imply $\Omav=1$ for all $\xi$. This is an invariant set associated with the  so--called ``parabolic'' solutions of (\ref{ellsq_dust}) and corresponding to $\kav=\RRav=K=0$ (see \cite{HM,HMM}). A full closed analytic solution (compatible with (\ref{funcs1})) is given by:
\bse\label{S1_ana2}\ba\Dm(\xi,r) &=&
\frac{\Dim}{[1+\Dim]\,\textrm{e}^{3\xi/2}-\Dim},\\
\HH(\xi,r) &=& \HH_i\,\textrm{e}^{-3\xi/2},\\
S(\xi,r) &=& -\frac{1}{2}\,\Dm(\xi,r).\ea\ese
so that solution curves remain for all $\xi$ in the line $S=-\Dm/2$ that lies in the plane $\Omav=1$ (see figure \ref{phase_space}). We have all variables fully determined. It is straightforward to verify that
(\ref{S1_ana2}) are fully compatible and equivalent to analytic
solutions given in terms of $\ell(ct,r)$ (see \cite{SG} and Appendix C). 

Critical points on this case are:
\begin{equation}{\bf C_1}:\quad \textrm{source},\qquad {\bf C_3}:\quad \textrm{sink},\end{equation}
The phase space evolution goes along the line $[S,-2S,1]$, from ${\bf C_1}$ to ${\bf C_3}$ (respectively, past and future global attractors). Since the critical point ${\bf C_3}$ is characterized by conditions of homogeneity $\Dm=S=0$, it corresponds to the Einstein--de Sitter  universe (dust FLRW model with zero spacial curvature). 

\subsection{FLRW dust models}

If  $\Dm=S=0$ with unrestricted $\Omav$, we have an invariant set associated with the particular homogeneous and isotropic sub--case of dust FLRW models. The critical points are in this case:
\begin{equation}{\bf C_3}:\quad \textrm{source},\qquad {\bf C_6}:\quad \textrm{sink},\end{equation}
Depending on the sign of $\Omiav-1$, the phase space evolution of these models will take place in the line $[0,0,\Omav]$. If $\Omiav-1<0$ the evolution goes between the source  ${\bf C_3}$ and the sink ${\bf C_6}$ (past and future attractors). If  $\Omiav-1>0$, it starts at ${\bf C_3}$ and goes upwards to $\Omav\to\infty$ towards maximal expansion. In the collapsing stage, the point ${\bf C_3}$ acts as a sink. If $\Omav=1$, we have the Einstein--de Sitter universe and the evolution is constrained to the point ${\bf C_3}$ with coordinates $[0,0,1]$. Also, we can identify the sink ${\bf C_6}$, whose coordinates are $[0,0,0]$, as Minkowski space--time in the Milne representation.

\subsection{RSC's} 

The solution curve corresponding to a RSC is also characterized by $\Dm=S=0$ with varying $\Omav$, thus the phase space evolution of that particular curve will also be along the line $[0,0,\Omav]$, just like the FLRW sub--case. This is not surprising because a RSC is a privileged isotropic observer. 

The fact that the RSC's have a separate evolution from the rest of the solution curves prevents the critical point ${\bf C_1}$ (a past and future attractor associated with the initial and collapsing singularities) to be global in all cases. However, ${\bf C_1}$ is a global attractor for all ``off center'' curves in configurations with RSC and for all curves in configurations lacking RSC (wormhole topology and vacuum limit).  

\subsection{Vacuum case}

It is known that the LTB metric (\ref{LTB1}) can describe the Schwarzchild-Kruskal space--time in terms of the world lines of its radial geodesic test observers with 4-velocity $u^a=\delta^a_0$. The equation of motion of these geodesics is identical to (\ref{Ysq_dust}) with $M=M_0$, so that $M'=0=\rho$ and  the Ricci tensor vanishes, while the constant $M_0$ can be identified with the ``Schwarzschild mass'' (see \cite{HM,HMM,Steph}). The function $\RR$ is the scalar 3-curvature of the hypersurfaces of simultaneity of these geodesic congruences and the average scalar curvature is related to their binding energy per unit rest mass of the test particle $E_i(r)=-K/2=(-1/12)\RRiav Y_i^2= (-1/2)H_0^2\kiav$. 

This particular case can also be constructed by assuming a Dirac delta distribution for $\rho_i$ so that a nonzero $\rhoiav$ follows from the integral definition (\ref{rhoave}). Also, without resorting to a Dirac delta, this vacuum case follows from the 3+1 formalism in equations (\ref{system2}) by setting $\rho=0$, so a definition of a nonzero average density follows from the Weyl tensor component in (\ref{EWeyl}) and (\ref{E}) as:
\begin{equation}  \EE = -\frac{\kappa c^2}{6}\,\rhoav=-\frac{\mav H_0^2}{2}=\frac{-M_0/Y_i^3}{\ell^3}\end{equation}
Since $m=\rho=0$ but $\mav>0$, as long as $\Gamma$ is finite (see section X) we have the following sufficient condition to characterize the Schwarzchild-Kruskal vacuum:
\begin{equation} \Dm =-1\quad \textrm{for all}\quad (\xi,r)\end{equation}
which defines an invariant set, since all solution curves with initial condition $\Dim=-1$ will necessarily be constrained to the plane $\Dm=-1$, parametrized by $[S,\Omav]$. The phase space evolution and critical points depends on the sign of $\Omiav$, which depends in turn on the binding energy of the radial geodesics. 

For zero binding energy $\kiav= 0=\Omiav- 1$ (the so--called Lema\^\i tre coordinates for Schwarzschild space-time~\cite{Steph}) the critical points are the global future and past attractors: 
\begin{equation}{\bf C_1}:\quad \textrm{source},\qquad {\bf C_2}:\quad \textrm{sink},\end{equation}
and phase space evolution is the line $[S,-1,1]$, from the source ${\bf C_1}$ to the sink ${\bf C_2}$.

For geodesics with negative binding energy: $\kiav>0$ or $\Omiav>1$ (the so--called Novikov coordinates~\cite{MTW}) the critical points are the same as those of the zero energy case, but ${\bf C_2}$ is now a saddle. The phase space evolution of the curves begin at the global past attractor (source) ${\bf C_1}$, approach the saddle ${\bf C_2}$ and go upwards towards diverging $\Omav$ at the maximal expansion, terminating again at the global future attractor (sink) ${\bf C_1}$.   

For geodesics with positive binding energy ($\kiav<0$ or $0<\Omiav<1$) the critical points are:
\ba &{\bf C_1}:&\quad \textrm{source},\qquad {\bf C_5}:\quad \textrm{sink},\nonumber\\
&{\bf C_2}:&\quad \textrm{saddle},\qquad {\bf C_4}:\quad \textrm{saddle}.\ea
In this case the curves begin at the global past attractor (source) ${\bf C_1}$, terminate at the global future attractor (sink) ${\bf C_5}$, approaching the saddles ${\bf C_2}$ and ${\bf C_4}$. This case is examined in subsection F of next section (see figure \ref{vacumcase}).           

We examine the phase space evolution of solution curves of several general case dust LTB configurations in the following section.


\section{LTB dust configurations}

\subsection{Expanding low density universe} 

A low density perpetually expanding dust configuration follows by selecting $\miav(0) <1$ and negative $\kiav$ for all $r$ defined in an initial hypersurface  $\T_i$ with an open topology having a RSC. This leads to $\Omiav<1$ or to ``hyperbolic'' dynamics. Such a configuration can be realized with the following initial value functions
\ba m_i &=& \frac{0.9}{1+\tan^2\,r},\nonumber\\
    k_i &=&  -\frac{\tan^2\,r}{1+\tan^2\,r},\nonumber\\
    Y_i &=&  H_0^{-1}\,\tan\,r,\nonumber\\
    \label{lowdens_clump_ivf}\ea
so that we have an initial density clump ($-1<\Dim\leq 0$) but a curvature void ($0\leq \Dik <2/3$). It is easy to show that these initial conditions yield  $\Omiav <1$ for all $r$, so that $0<\Omav<1$ for all $\xi$. These conditions also comply with the no--shell--crossing conditions (\ref{nxcr_ini}), so dust layers monotonously expand perpetually from an initial big bang. 

\begin{figure}[htbp]
\begin{center}
\includegraphics[width=2.5in]{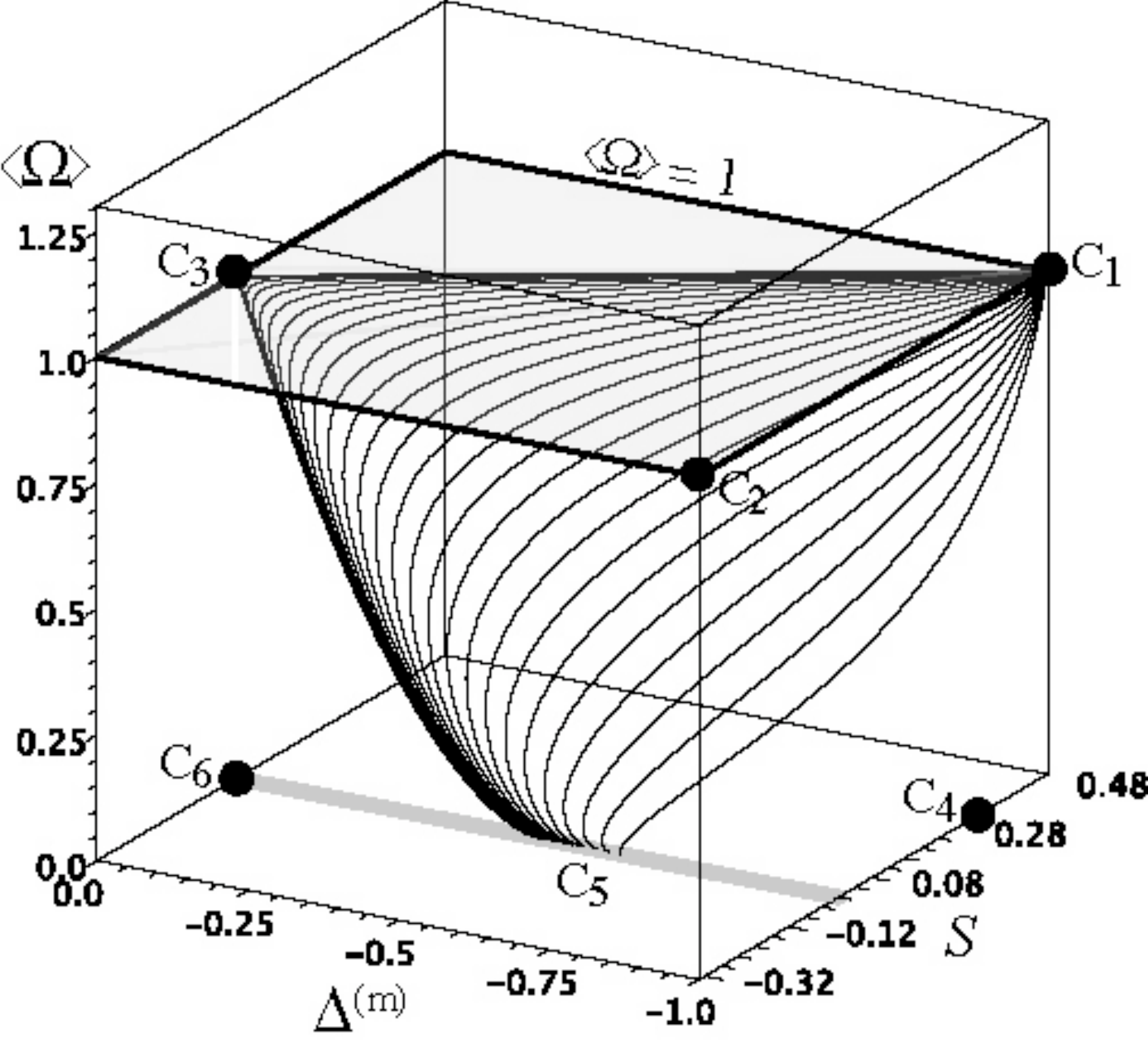}
\caption{{\bf Low density configuration.} Dust layers expand from the past attractor (source) ${\bf C_1}$, denoting the big bang singularity, approaches the saddle ${\bf C_3}$ and drops towards the future attractor given by the line of sinks ${\bf C_5}$. Notice that (\ref{lowdens_clump_ivf}) implies that the evolution of the RSC is constrained to ${\bf C_3}$ for all time. }
\label{lowdens_clumpPS}
\end{center}
\end{figure}

As shown by figure \ref{lowdens_clumpPS} the solution curves start at the past attractor or source ${\bf C_1}$ (big bang), evolve towards the saddle ${\bf C_3}$ and terminate at the future attractor given by the line of sinks ${\bf C_5}$ in the plane $\Omav=0$. The evolution of the RSC is constrained to ${\bf C_3}$ for all time. The function $\Dm$ remains negative for all $\xi$, thus we have density clumps at all $\T$.   
%
%

\subsection{Low density void formation}

An interesting variation of a perpetually expanding configuration, similar to the one presented in subsection A and figure \ref{lowdens_clumpPS}, is that where an initial density clump ($-1<\Dim<0$) evolves for all curves into density voids with $\Dm>0$. In this subsection we provide a numeric example of dust configurations discussed in reference \cite{mustapha} (see also \cite{bon_ch,BKH}).

\begin{figure}[htbp]
\begin{center}
\includegraphics[width=2.5in]{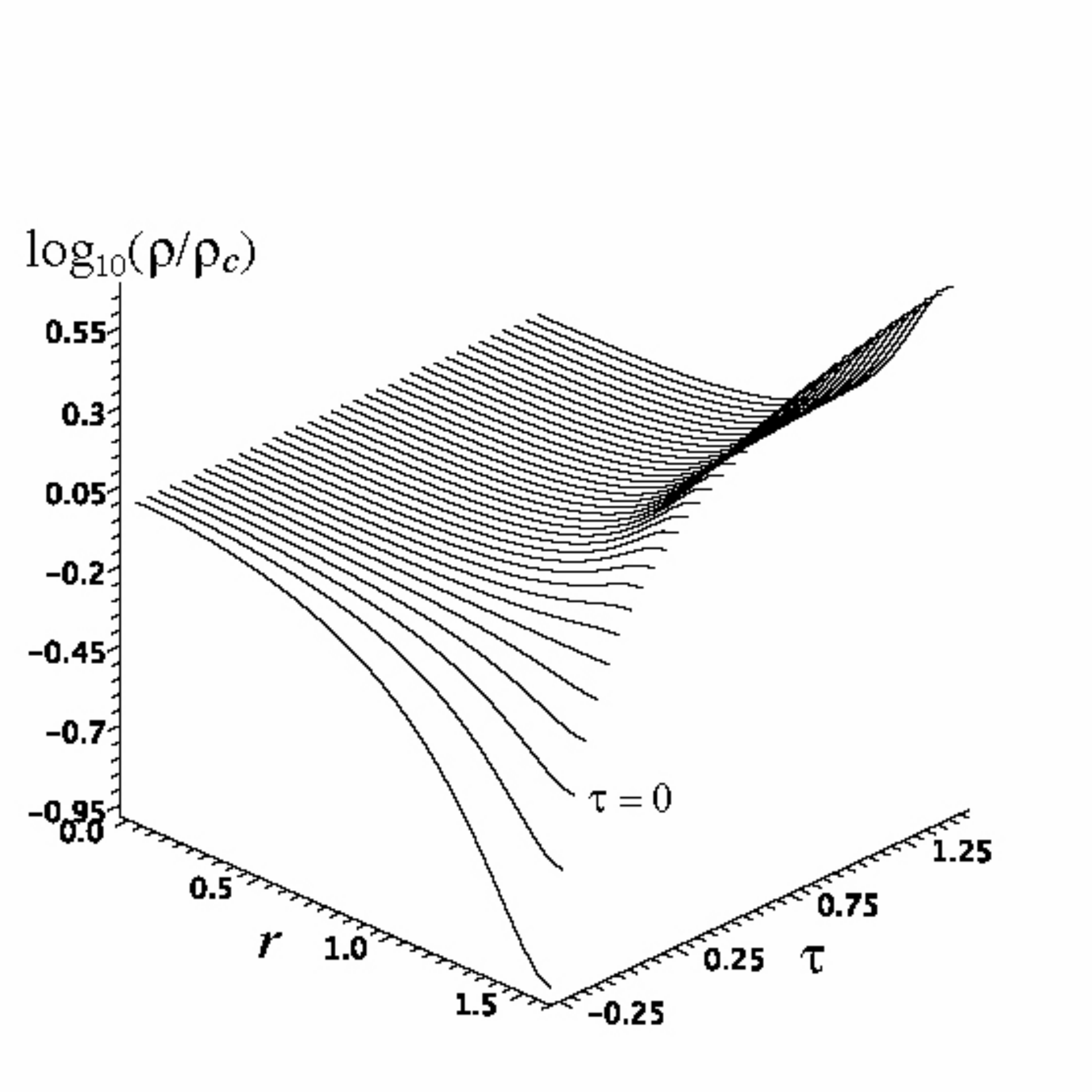}
\caption{{\bf Low density voids: density profiles} The picture depicts the normalized radial density profile $\rho/\rho_c$ for different hypersurfaces $\T$ marked by constant $\tau$. Notice how an initial clump at $\tau=0$ evolves into a voids  profile for latter $\tau$.}
\label{dens_profs}
\end{center}
\end{figure}

An initial dust clump transforming into a void emerges by choosing the following initial value functions
\ba m_i &=& m_{01}+\frac{m_{00}}{1+\alpha_0^2\,\tan^2\,r},\nonumber\\ &{}& m_{00}= 0.05,\quad m_{01}=0.9,\quad \alpha_0=2.0,\nonumber\\
    k_i &=&  k_{01}+\frac{k_{00}}{1+\beta_0^2\,\tan^2\,r},\nonumber\\ &{}& k_{00}=-5.0,\quad k_{01}=-1.0,\quad \beta_0=0.5,\nonumber\\
    Y_i &=&  H_0^{-1}\,\tan\,r,\nonumber\\
    \label{lowdens_void_ivf}\ea
which comply with the no--shell--crossing conditions (\ref{nxcr_ini}).
The evolution of this configuration is similar to the previous one: dust layers also expand perpetually (``hyperbolic'' dynamics) from an initial big bang and the hypersurfaces $\T$ have an open topology with a RSC. However, the contrast function, $\Dm$, now passes from negative to positive as $\xi$ increases, indicating that initial clumps transform into voids. This can be seen very clearly if we plot (see figure (\ref{dens_profs})) the normalized density profiles along the hypersurfaces $\T$ from the function $\rho/\rho_c=m/m_c=\mav (1+\Dm)/m_c$, where the subindex ${}_c$ denotes evaluation along the RSC marked by $r=0$. The density as function of $\tau$ and $r$ can be readily found by solving numerically the system (\ref{sys3}) for initial conditions (\ref{lowdens_void_ivf}). This transition from clumps to voids is depicted by figure \ref{dens_profs}. 
%
%

\begin{figure}[htbp]
\begin{center}
\includegraphics[width=2.5in]{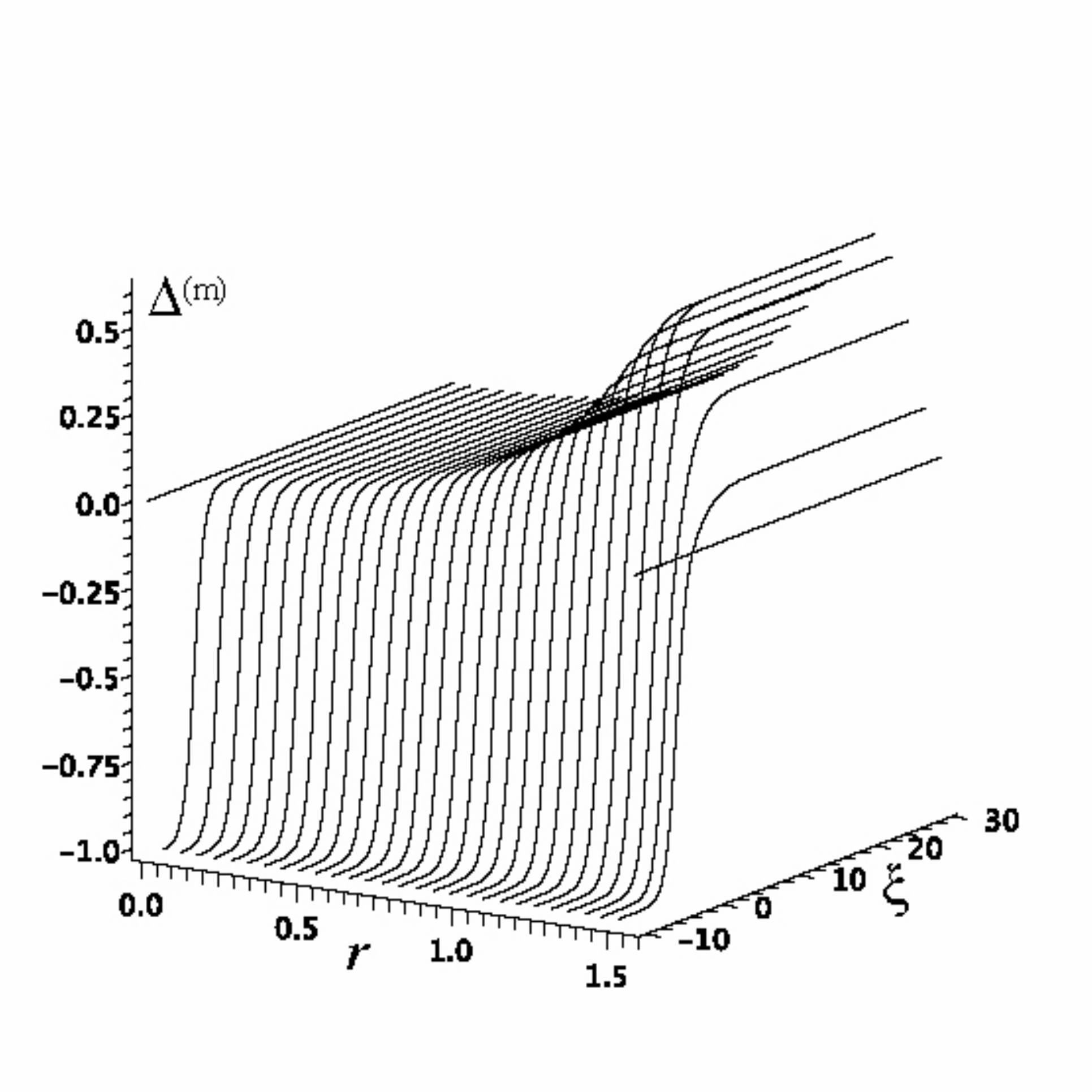}
\caption{{\bf Density contrast function} The picture depicts the function $\Dm$ passing for all $0<r<\pi/2$ from negative to positive indicating that an initial clump evolves into a void  profile. Notice how $\Dm=0$ at the RSC $r=0$, and asymptotically at $r\to\pi/2$ (or $Y_i\to\infty$). For solution curves with $r$ close to the RSC, $\Dm$ remains close to zero but is nevertheless positive.}
\label{lowdens_void_delta}
\end{center}
\end{figure}

This transition from clumps to voids can also be seen from the form of the contrast density function $\Dm$ given in terms of $\xi$ and $r$, displayed in figure \ref{lowdens_void_delta} which shows how $\Dm$ is initially negative but becomes positive for all solution curves with $r$ constant (except the RSC at $r=0$) as $\xi$ increases.
%
%

\begin{figure}[htbp]
\begin{center}
\includegraphics[width=2.5in]{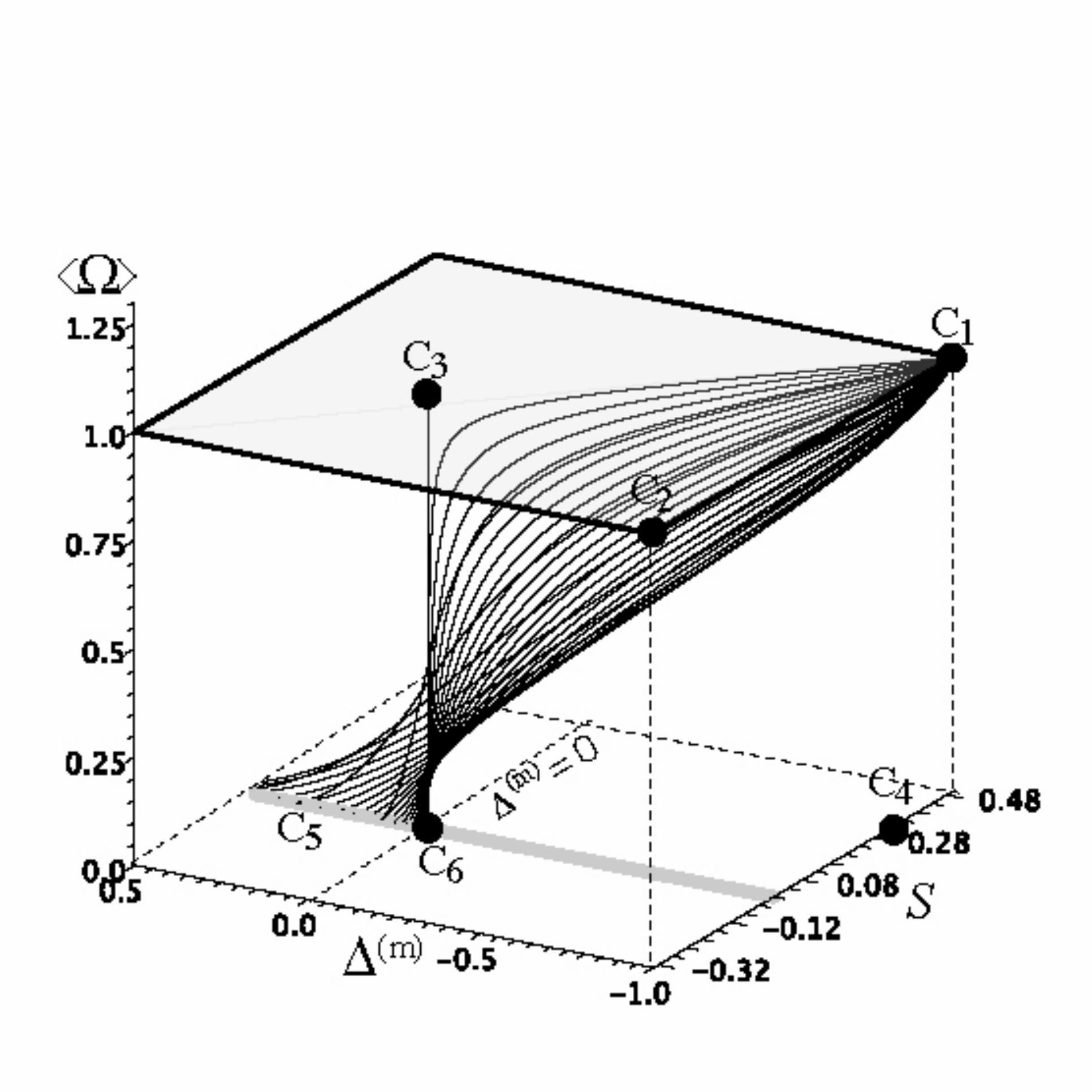}
\caption{{\bf Low density voids} The evolution is similar to that of Fig 1, except that $\Dm$ becomes positive as all  solution curves drop to the line of sinks $C_5$.}
\label{lowdens_void}
\end{center}
\end{figure}

The evolution of this configuration in phase space is shown by figure \ref{lowdens_void}. Solution curves start at the past attractor (source) ${\bf C_1}$ (big bang), approach the saddle ${\bf C_3}$ and evolve towards the future attractor (line of sinks) ${\bf C_5}$ in the plane $\Omav=0$. The RSC evolves from the source ${\bf C_3}$ towards the sink ${\bf C_6}$. However, now the curves evolve in such a away that all curves (save the RSC) hit the line of sinks ${\bf C_5}$ at positive values of $\Dm$ (passage from clumps to voids). 
%
%


\subsection{Re--collapsing high density universes}

Re--collapsing high density configurations follow by selecting $\miav(0)>1$ and $\kiav>0$ for all $r$, leading to $\Omiav \geq 1$ for all $r$, thus $\Omav \geq 1$ for all $\xi$.  Dust layers expand from an initial big bang, reach a maximal expansion and then collapse (the so--called ``elliptic'' dynamics~\cite{SG,HM,HMM}). These configurations are compatible with $\T_i$ having either a closed topology with two RSC, an open topology with one RSC or a wormhole topology without RSC (see Appendix A). We examine the closed and open case separately below, and the wormhole case in section E.  

\begin{figure}[htbp]
\begin{center}
\includegraphics[width=2.5in]{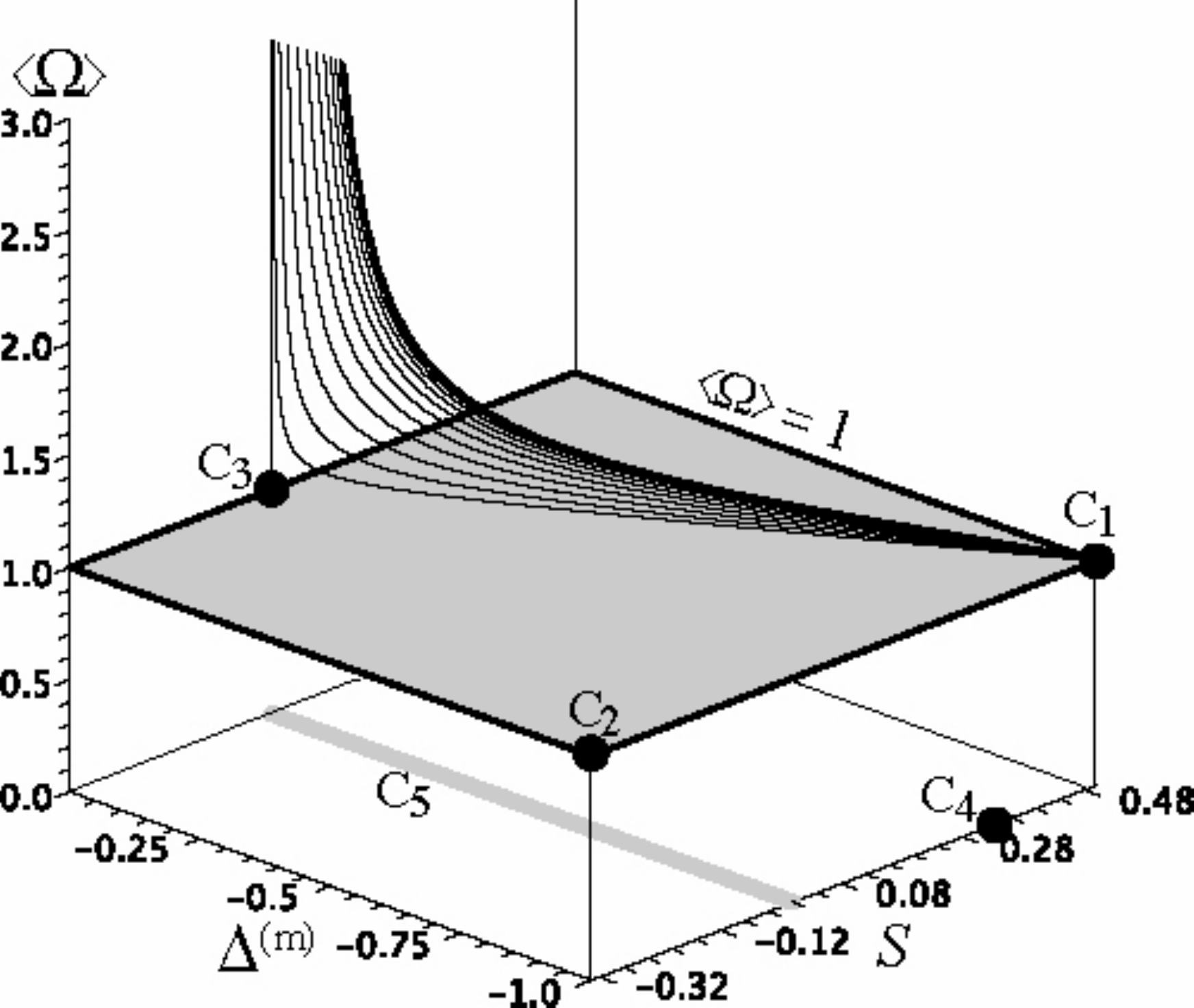}
\caption{{\bf Phase space evolution of a high density re--collapsing universe with closed topology}. Solution curves evolve from the past attractor or source ${\bf C_1}$, approach the saddle ${\bf C_3}$ and diverge as $\xi\to \ln[\Omiav/(\Omiav-1)]$, marking the maximal expansion of dust layers. Only the range $0\leq r\leq \pi/2$ is shown, with the RSC at $r=0$ evolving from the source ${\bf C_3}$ upwards. The off--center layers approaching the saddle $C_3$ are those with $r\approx 0$ while those further away from ${\bf C_3}$ are close to the ``turning value'' $r=\pi/2$. Layers marked by $\pi/2\leq r\leq \pi$, including the second RSC at $r=\pi$, would have identical evolution as those shown in the figure, with those marked by $r\approx \pi$ near the saddle ${\bf C_3}$. }
\label{spheric_top}
\end{center}
\end{figure}

The case with spherical ``closed'' topology can be obtained with the following initial value functions:
\ba m_i &=& m_{01}+\frac{m_{00}-m_{01}}{1+\sin^2\,r},\qquad m_{00}=3.9,\quad m_{01}=1.1\nonumber\\
    k_i &=&  k_{01}+\frac{k_{00}-k_{01}}{1+\sin^2\,r},\qquad k_{00}=2.9,\quad k_{01}=0.22\nonumber\\
    Y_i &=&  H_0^{-1}\,\sin\,r,\nonumber\\
    \label{spheric_top_ivf}\ea
which comply with (\ref{TPcond}) at the turning point $r=\pi/2$. 
The phase space evolution is illustrated in figure \ref{spheric_top}. The solution curves start at the past attractor (source) ${\bf C_1}$ (big bang) and evolve upwards as $\Omav\to \infty$, which corresponds to maximal expansion given by $\HH\to 0$ and $\xi\to \ln[\Omiav/(\Omiav-1)]$. The RSC also evolves upwards from the source ${\bf C_3}$. For off center solution curves this critical point is a saddle.
%
%

\begin{figure}[htbp]
\begin{center}
\includegraphics[width=2.5in]{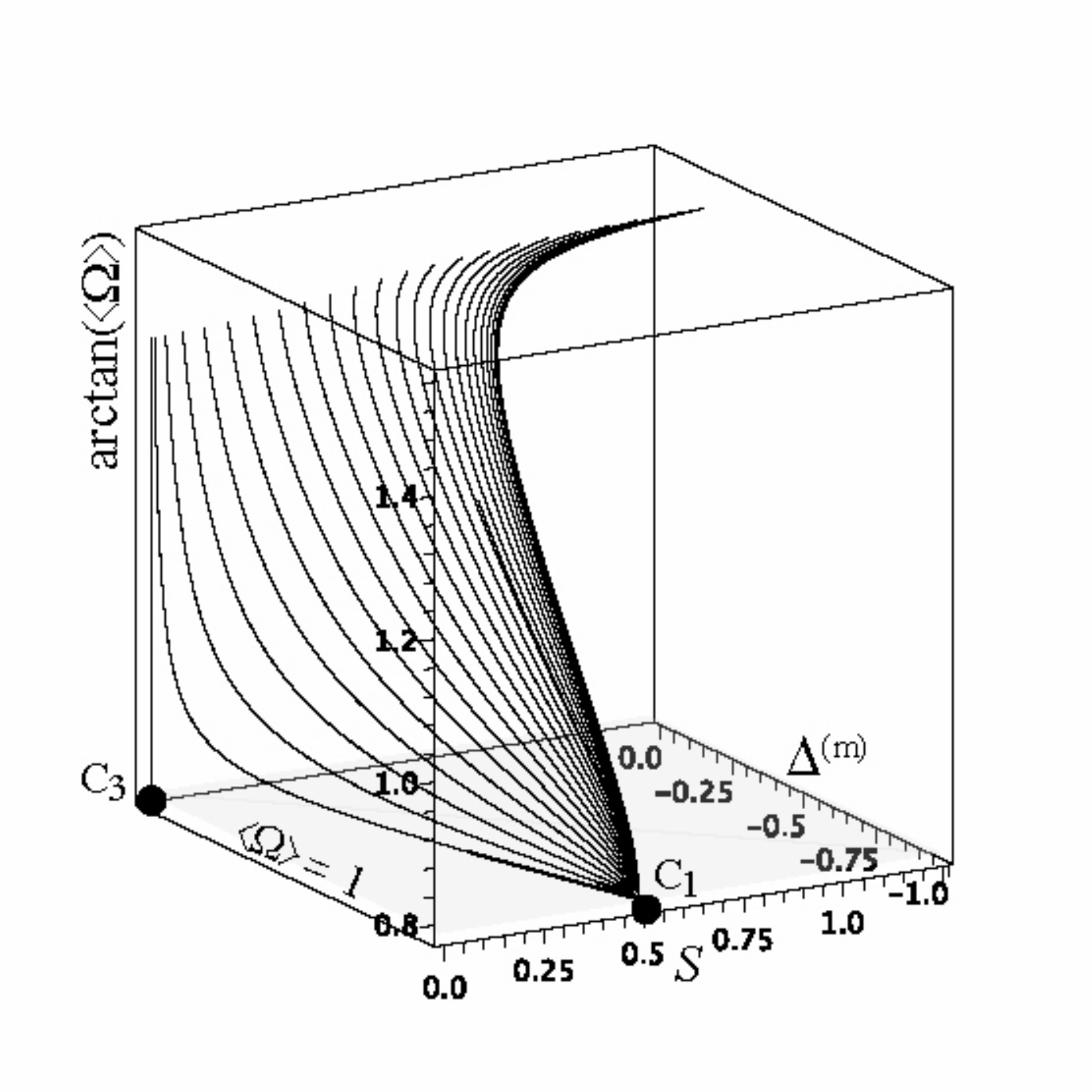}
\caption{{\bf Phase space evolution of the collapsing stage in the re--collapsing universe with closed topology}. Only the end stage of the collapse is described by plotting the phase space variables by means of system (\ref{sys3}). Solution curves evolve from infinite values of $\Omav$ and $S$ to the point ${\bf C_1}$, which is now a future attractor or sink. The curves also approach the saddle ${\bf C_3}$. Only the range $0\leq r\leq \pi/2$ is shown, with the RSC at $r=0$ evolving from infinite values of $\Omav$ to the ${\bf C_3}$, which is now a sink. Layers marked by $\pi/2\leq r\leq \pi$, including the second RSC at $r=\pi$, would have identical evolution as those shown in the figure, with those marked by $r\approx \pi$ near the saddle ${\bf C_3}$. }
\label{collapse}
\end{center}
\end{figure}

Since the collapsing stage cannot be described with solution curves of system (\ref{system5}), we can examine this stage with system (\ref{sys3}) by plotting the curves $[S(\tau,r),\Dm(\tau,r),\Omav(\tau,r)]$, where $S$ and $\Omav$ are defined by (\ref{S_def}) and (\ref{Omega_def}). As shown by figure \ref{collapse}, the curves come downwards from infinite $\Omav$ and $S$ to the critical point ${\bf C_1}$, which is now a future attractor or sink, associated with a second curvature singularity (``big crunch''). As in the expanding state, curves near the RSC approach the saddle ${\bf C_3}$, while the RSC evolves along the line $[0,0,\Omav]$ from infinite $\Omav$ to $\Omav =1$ at the sink ${\bf C_3}$.  
%
%

The case with open topology with one RSC can be obtained with the following initial value functions:
\ba m_i &=& \frac{5.0}{1+\tan^2\,r},\nonumber\\
    k_i &=& \frac{2.0}{1+\tan^2\,r},\nonumber\\
    Y_i &=&  H_0^{-1}\,\tan\,r,\nonumber\\
    \label{closed_tan_ivf}\ea
As in the case with closed topology, dust layers emerge from a big bang singularity reach a maximal expansion and then re-collapse in a big crunch, but now dust layers far from the RSC at $r=0$ re-collapse in very long time which becomes infinite in the limit $r\to\pi/2$. 

\begin{figure}[htbp]
\begin{center}
\includegraphics[width=2.5in]{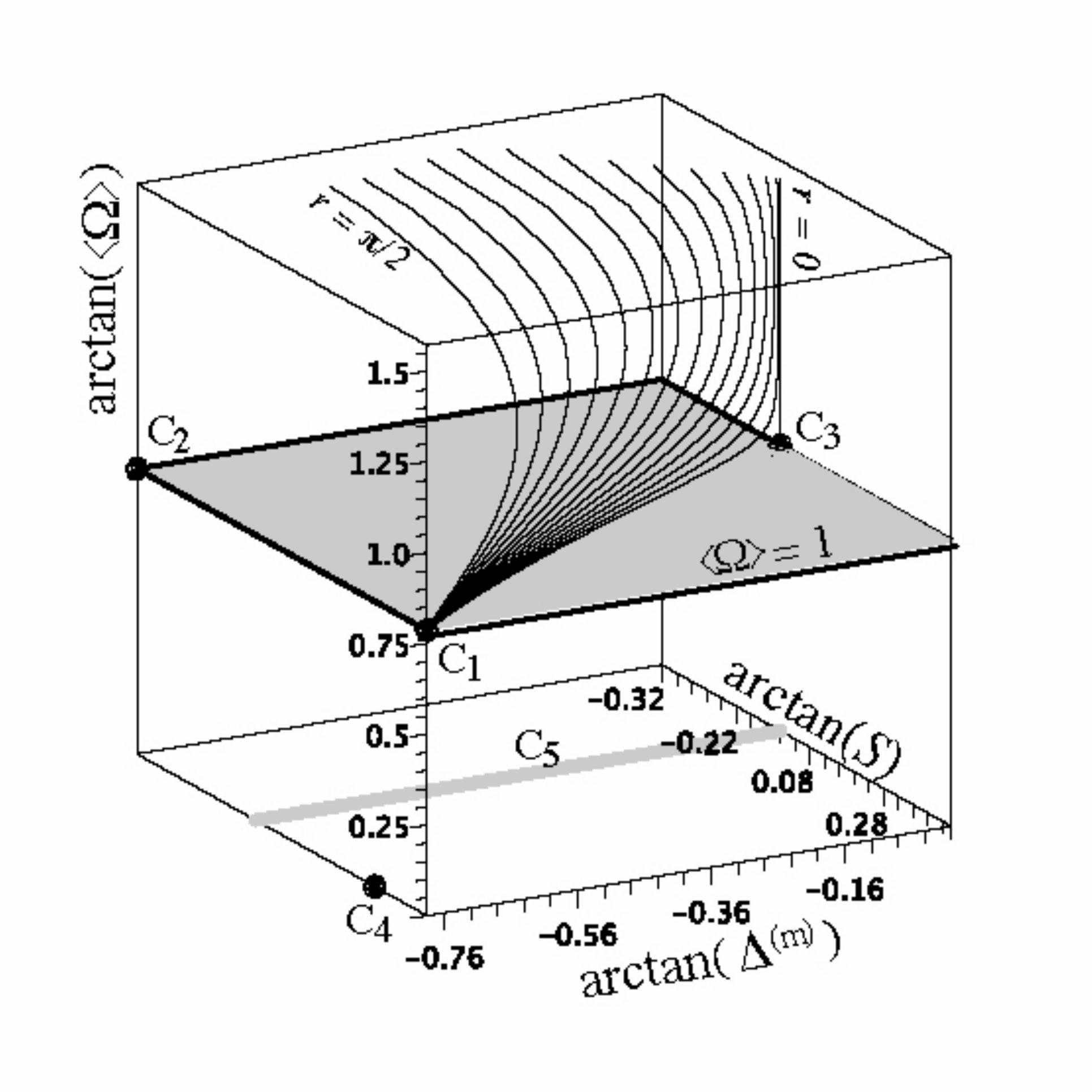}
\caption{{\bf Phase space evolution of high density re--collapsing universe with open topology.} Since $\Omav\to\infty$ as $\xi\to\Omiav/(\Omiav-1)$, we plot $\arctan\,\Omav$, The curves start at the past attractor or source ${\bf C_1}$ (big bang) approach the saddle ${\bf C_3}$ and diverge upwards (maximal expansion). The RSC evolves from the source ${\bf C_3}$ to $\Omav\to\infty$. The collapsing stage is not described. }
\label{closed_tan}
\end{center}
\end{figure}

The evolution of the solution curves of system (\ref{system5}) in phase space is shown by figure \ref{closed_tan}. This evolution is similar to the closed spherical case depicted in figure \ref{spheric_top}. Since initial conditions yield $\Omiav \geq 1$ we have $\Omav \geq 1$ for all $\xi$. The curves (except the RSC) start at the past attractor or source ${\bf C_1}$ (big bang), approach the saddle ${\bf C_3}$ and diverge upwards as $\xi\to\ln[\Omiav/(\Omiav-1)]$, corresponding to maximal expansion ($\HH\to 0$). The RSC evolves from the source ${\bf C_3}$ towards $\Omav\to\infty$.  
%
%

\begin{figure}[htbp]
\begin{center}
\includegraphics[width=2.5in]{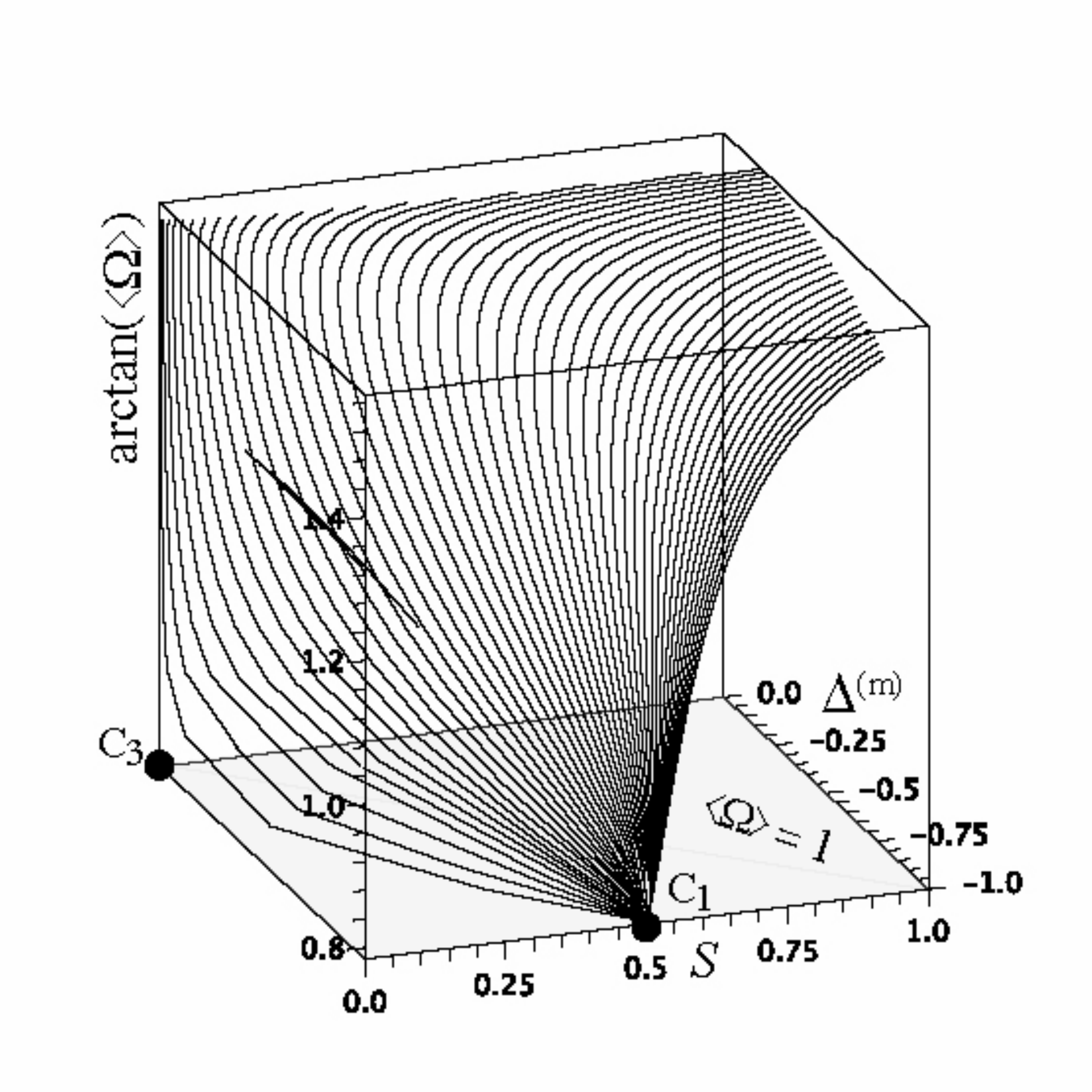}
\caption{{\bf Phase space evolution of the collapsing stage in the re--collapsing universe with open topology}. Phase space variables are plotted by solving system (\ref{sys3}). Solution curves evolve from infinite values of $\Omav$ and $S$ to the point ${\bf C_1}$, which is now a future attractor or sink. The RSC at $r=0$ evolves from infinite values of $\Omav$ to the sink ${\bf C_3}$. }
\label{collapse_ct}
\end{center}
\end{figure}

As in the case with spherical topology, the collapsing stage cannot described with (\ref{system5}). We use then the solution curves of (\ref{sys3}) to plot the phase space variables in the collapsing stage in figure \ref{collapse_ct}, which is very similar to figure \ref{collapse}: the critical point ${\bf C_1}$ is now a future attractor or sink associated with the collapsing singularity (``big crunch''), while curves near the RSC approach the saddle ${\bf C_3}$ and the RSC evolves from infinite $\Omav$ downwards to $\Omav=1$ along the line $[0,0,\Omav]$.
%
%


\subsection{Structure formation scenario}

A very important type of configuration, associated with structure formation scenarios, follows by having a local region around a RSC evolving from a big bang singularity and collapsing into a black hole (second singularity), while the region ``outside'' expands perpetually as a sort of ``cosmic background''. This type of configurations have also been examined with the LTB original variables (see \cite{kras, HM,HK1,HK2,HK3,HK4} and references quoted therein).

The structure formation scenario requires that $\Omiav-1$ be positive at the RSC and becomes negative for large $r$. It can be constructed by choosing $m_i>1$ near the RSC that becomes $m_i<1$ asymptotically, while $k_i$ must be positive at the RSC and become asymptotically negative. Such a choice will yield for re--collapsing curves $\Omiav>1$ near the RSC, so that $\Omav\leq 1$ and diverges for $\xi\to \ln[\Omiav/(\Omiav-1)]$. For expanding curves we have $0<\Omiav\leq 1$ and $0<\Omav\leq 1$ for all $\xi$. The following initial value functions fulfill these conditions:
\ba m_i &=& m_{01}+\frac{m_{00}-m_{01}}{1+r^6},\qquad m_{00}=1.5,\quad m_{01}=0.2\nonumber\\
    k_i &=&  k_{01}+\frac{k_{00}-k_{01}}{1+r^6},\qquad k_{00}=0.5,\quad k_{01}=-0.5\nonumber\\
    Y_i &=&  H_0^{-1}\,r,\nonumber\\
    \label{struct_form_ivf}\ea

We have depicted in all the graphs the collapsing solution curves in red and the perpetually expanding ones in blue. The collapsing curves evolve as the solution curves of figures  \ref{spheric_top} and \ref{closed_tan}, except that now these curves only cover a bounded range of $r$ around the RSC at $r=0$, with expanding curves filling the asymptotic range of $r$. 

\begin{figure}[htbp]
\begin{center}
\includegraphics[width=2.5in]{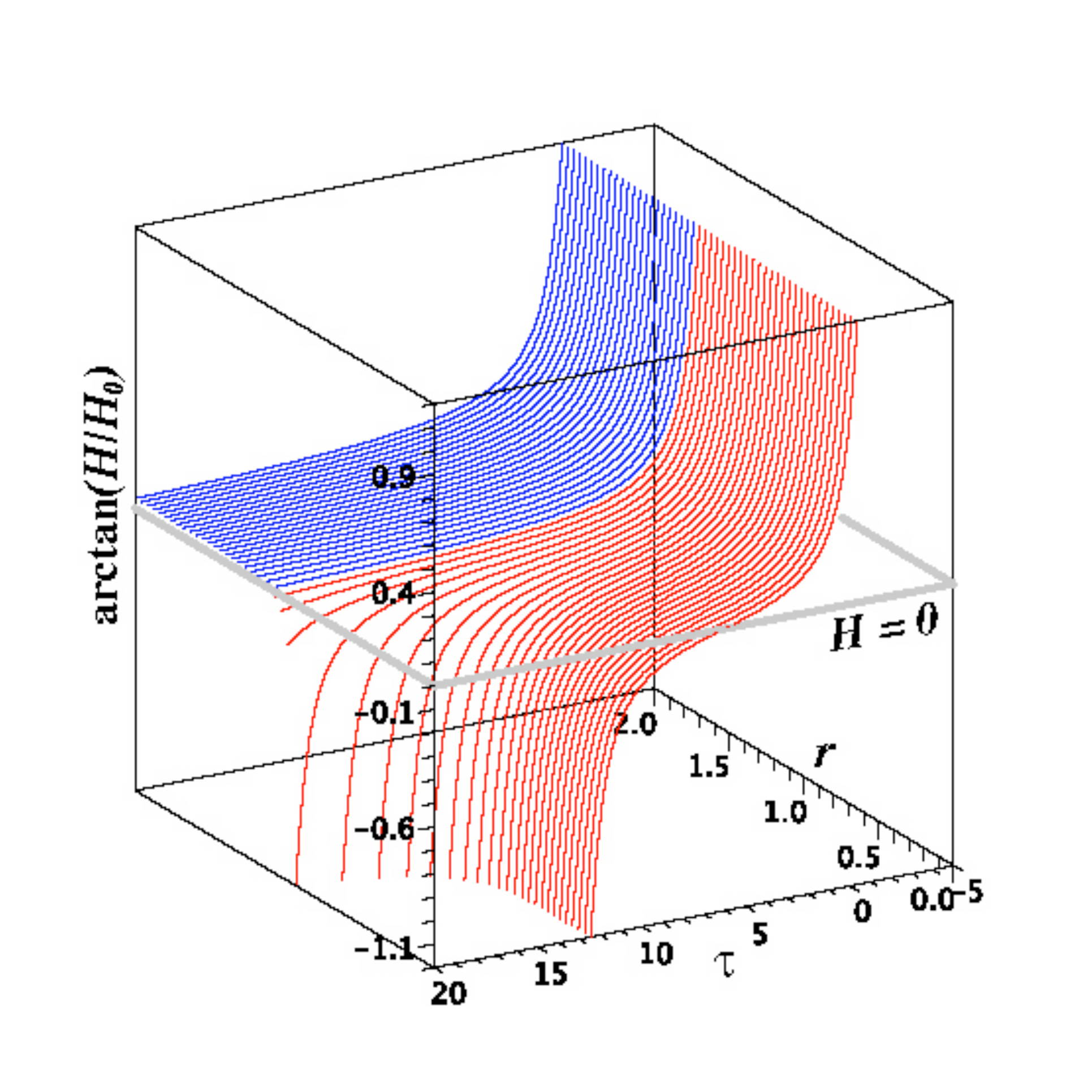}
\caption{{\bf The function $\HH$ in a structure formation scenario.} This function is obtained by solving the system (\ref{sys3}) for initial conditions (\ref{struct_form_ivf}). For re--collapsing curves with $\Omiav>1$ (red color) this function passes from positive (expanding) to negative (collapsing), while for perpetually expanding curves with $\Omiav\leq 1$ (blue) it remains always positive.}
\label{struct_formH}
\end{center}
\end{figure}

For the collapsing curves the function $\HH$ obtained by solving the system (\ref{sys3}) for initial conditions (\ref{struct_form_ivf}) changes sign (passes from positive to negative). This function is plotted in figure \ref{struct_formH} showing collapsing and perpetually expanding layers respectively in red and blue colors.
%
%

\begin{figure}[htbp]
\begin{center}
\includegraphics[width=2.5in]{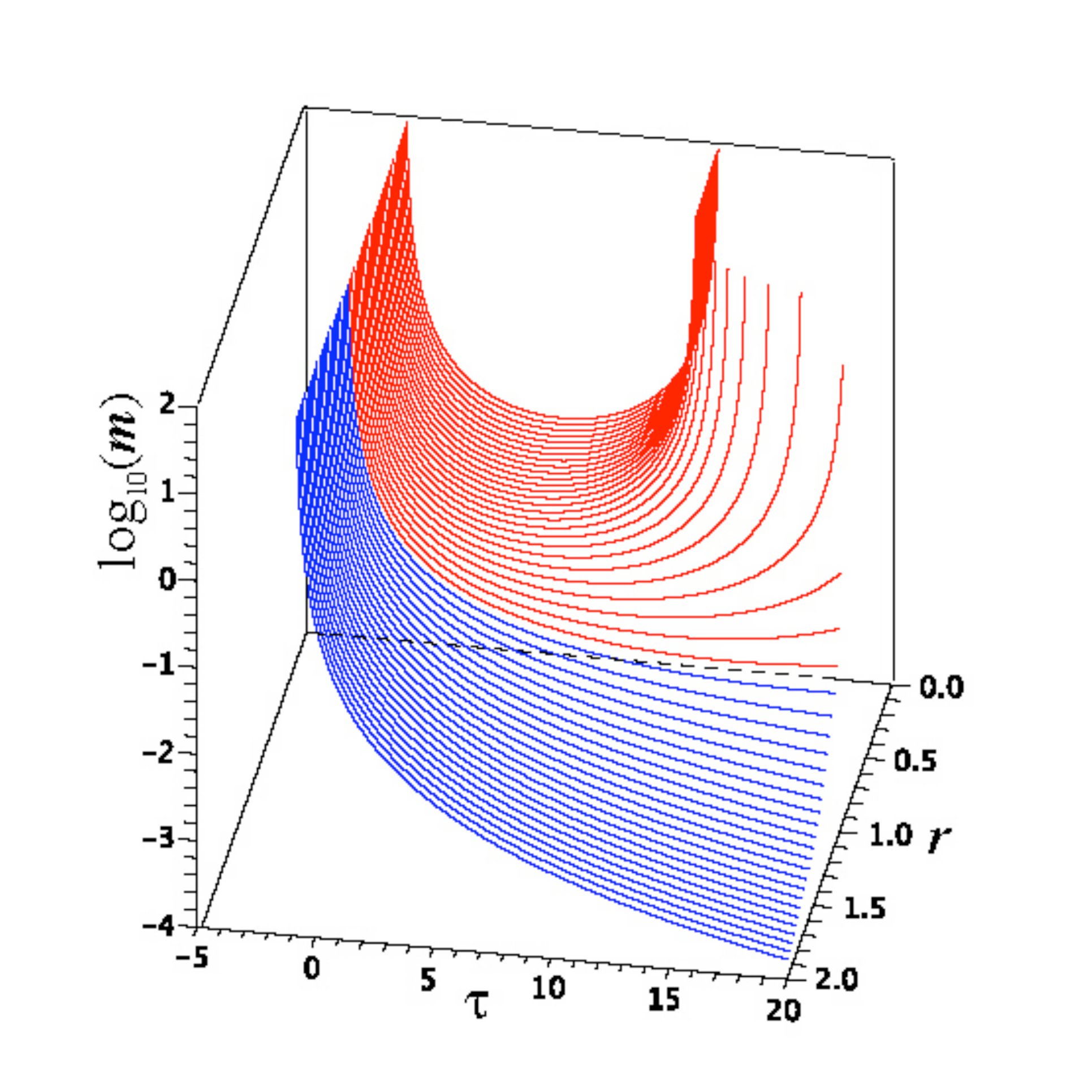}
\caption{{\bf Density for a structure formation scenario.} The initial big bang singularity is now associated to infinite density that decays as dust layers expand. Curves (red) with $\Omiav>1$ reach a minimal density (at maximal expansion) and then density rebounds to infinity at the collapse. For the remaining curves with $\Omiav\leq1$ (blue) density just dilutes monotonously as $\tau$ grows.}
\label{struct_form_rho}
\end{center}
\end{figure}

It is also interesting to visualize the the dimensionless density $m$ (see figure \ref{struct_form_rho}). For collapsing curves (red) around the RSC at $r=0$ this function decreases from infinite values as dust layers expand from the initial singularity, reaches a minimal value (maximal expansion) and then increases again as curves go into a collapsing singularity. However, for perpetually expanding curves (blue) this function decreases monotonically from the initial big bang. 
%
%

\begin{figure}[htbp]
\begin{center}
\includegraphics[width=2.5in]{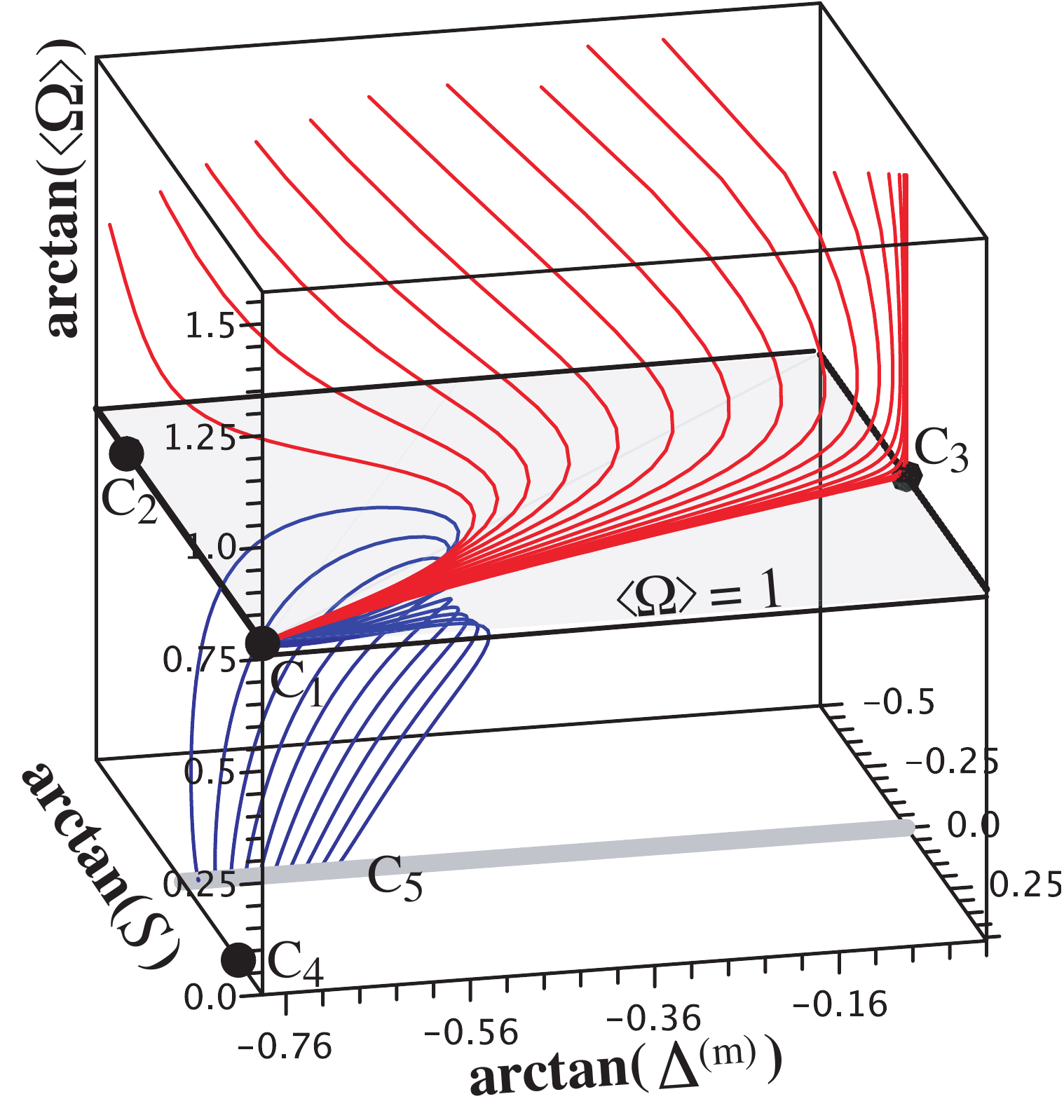}
\caption{{\bf Phase space picture of the structure formation scenario.} Red color corresponds to collapsing curves with $\Omiav>1$, while perpetually expanding curves with $0<\Omiav < 1$ are depicted in blue. All ``off center'' curves start at the past attractor ${\bf C_1}$, perpetually expanding curves descend into the future attractor given by the line of sinks ${\bf C_3}$, collapsing curves approach the saddle ${\bf C_3}$ and go upwards to diverging $\Omav$ (maximal expansion before collapse). The saddle ${\bf C_2}$ splits expanding from collapsing curves. The curve with $\Omiav=1$ (not shown) would go from ${\bf C_1}$ towards this saddle. }
\label{struct_form2}
\end{center}
\end{figure}

\begin{figure}[htbp]
\begin{center}
\includegraphics[width=2.5in]{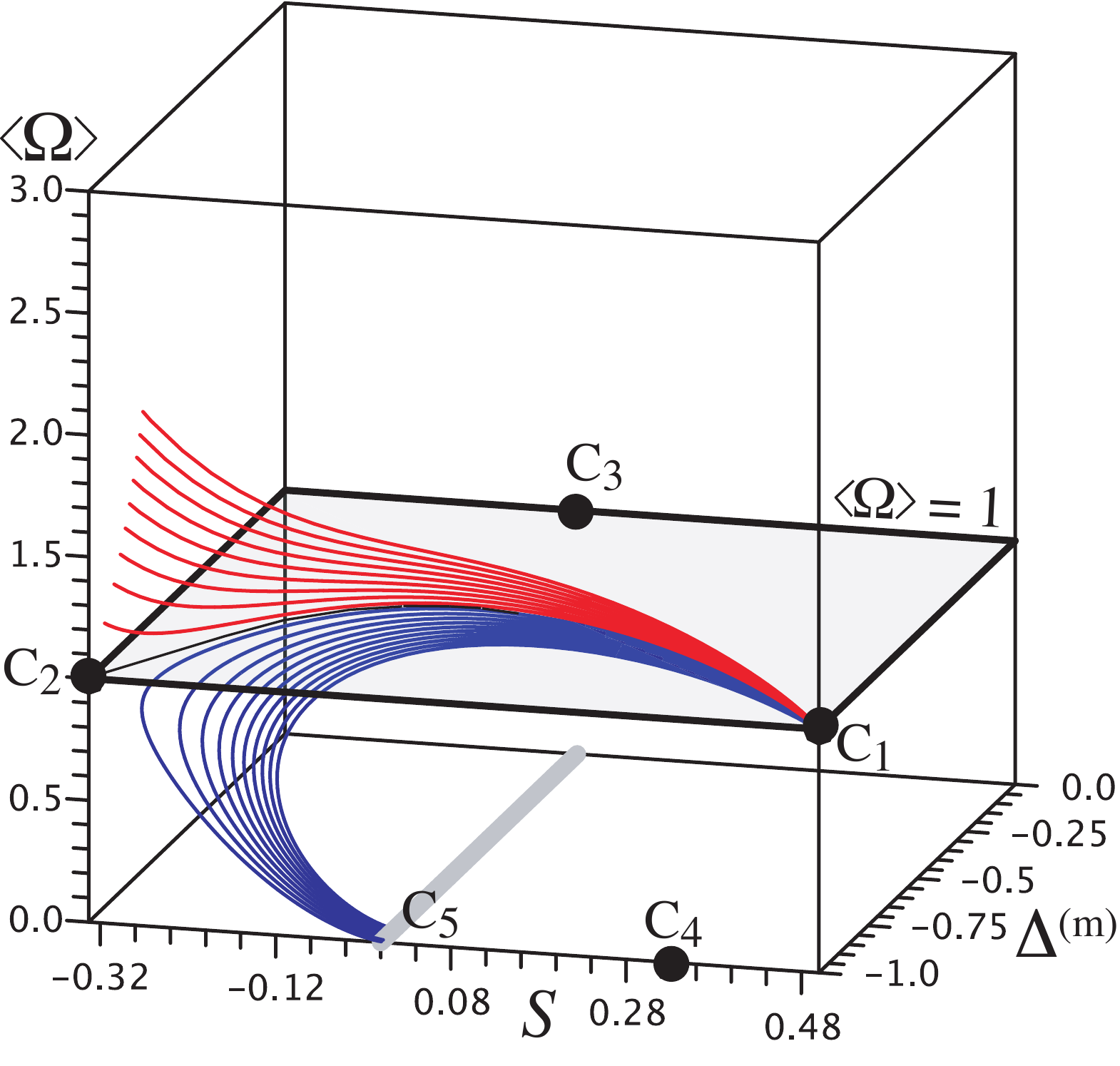}
\caption{{\bf Close up of the solution curves near the saddle $C_2$}.Red color corresponds to layers with $\Omiav>1$ and blue ones to $\Omiav<1$. The plot only depicts solution curves that correspond to $r$ values near the change from re--collapsing to perpetually expanding behavior (where $H$ becomes negative). Notice how this saddle splits collapsing curves going upwards to $\Omav\to\infty$ while expanding ones go downwards to line of sinks ${\bf C_5}$. Notice how the curve corresponding to $\Omiav=1$ remains in the plane $\Omav=1$, starting from the source $C_1$ and ending in the saddle $C_2$.}
\label{struct_form3}
\end{center}
\end{figure}

The evolution of the solution curves of (\ref{system5}) in phase space is more complicated than for previously presented configurations. Collapsing curves (red) with $\Omiav > 1$ evolve as curves in re--collapsing high density configurations of figures \ref{spheric_top} and \ref{closed_tan}, starting at the past attractor or source ${\bf C_1}$ (big bang), approaching the saddle ${\bf C_3}$ and diverging upwards ($\Omav\to\infty$) as $\xi\to\ln[\Omiav/(\Omiav-1)]$. Perpetually expanding curves (blue) with $0<\Omiav\leq 1$ resemble the curves of the low density universe of figure \ref{lowdens_clumpPS}, starting at the same past attractor or source ${\bf C_1}$ (big bang) approaching the saddle ${\bf C_3}$ and descending towards the future attractor given by line of sinks ${\bf C_5}$. This is shown in figures \ref{struct_form2} and \ref{struct_form3}. 

As shown by figure \ref{struct_form2}, all curves start at the past attractor ${\bf C_1}$ (big bang), those near the RSC ($r=0$) approach the saddle ${\bf C_3}$ (see figure \ref{struct_form2}) while those near the split from collapsing to expanding approach the saddle ${\bf C_2}$ (see figure \ref{struct_form3}). All collapsing curves end up diverging upwards $\Omav\to\infty$ as $\xi\to \ln[\Omiav/(\Omiav-1)]$ (maximal expansion). Blue curves with with $0<\Omiav\leq 1$ are perpetually expanding with $0<\Omiav\leq 1$ and so $0<\Omav\leq 1$ for all $\xi$. Expanding curves also start at the source ${\bf C_1}$ (big bang), but do not approach the saddle ${\bf C_3}$, instead they approach  the saddle ${\bf C_2}$ and end up in the line of sinks ${\bf C_5}$. The RSC evolves from the source ${\bf C_3}$ upwards. Both the collapsing and expanding curves approach the saddle ${\bf C_2}$, which splits the former curves going upwards to $\Omav\to\infty$ and the latter downwards towards the line of sinks ${\bf C_5}$. The curve with $\Omiav=1$ remains in the plane $\Omav=1$, evolving from ${\bf C_1}$ towards ${\bf C_2}$. See figure \ref{struct_form3}.

\begin{figure}[htbp]
\begin{center}
\includegraphics[width=2.5in]{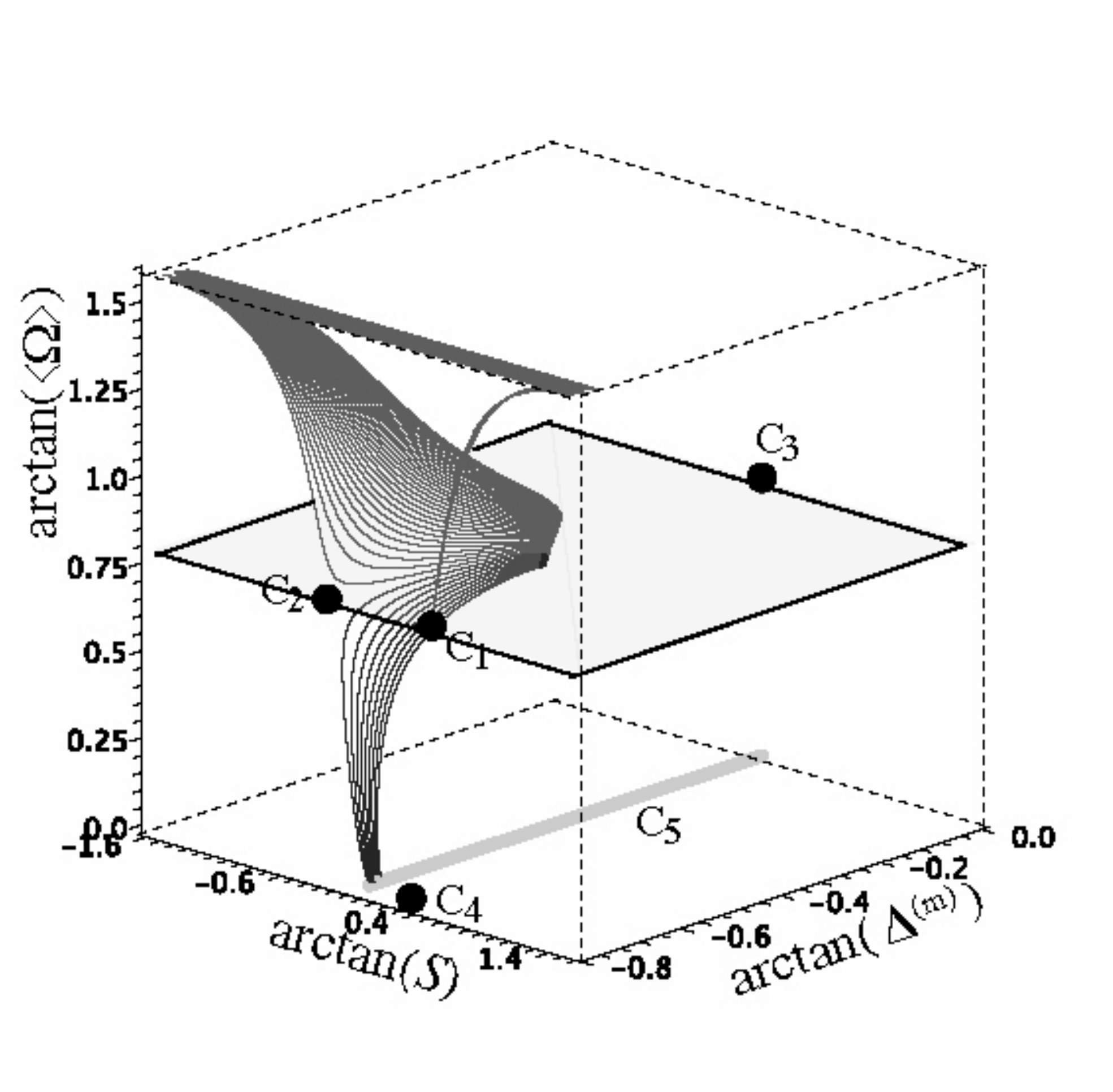}
\caption{{\bf Collapsing stage for curves of figure \ref{struct_form3}.} The curves come from solving (\ref{sys3}) for initial conditions (\ref{struct_form_ivf}). The end state of collapsing curves (red) is the point ${\bf C_1}$, which is now a sink. Collapsing curves go upwards taking infinite values of $\Omav$ and $S$. Notice how the saddle ${\bf C_2}$ splits collapsing curves going upwards to $\Omav\to\infty$ while expanding ones (blue) go downwards to the future attractor given by the sinks ${\bf C_5}$.}
\label{collapse_sf}
\end{center}
\end{figure}

The collapsing curves behave near the collapse as the curves in figures \ref{collapse} and \ref{collapse_ct}. We show in figure \ref{collapse_sf} the collapsing stage of the curves plotted in figure \ref{struct_form3}, for a range of $r$ around values where $\Omiav-1$ passes from positive (collapsing curves) to negative (expanding curves). Notice how collapsing curves end up falling into the critical point ${\bf C_1}$, which is a future attractor or sink. As the curves approach the splitting saddle ${\bf C_2}$, the collapsing ones go to infinite $\Omav$ and $S$, while expanding ones fall into the sinks ${\bf C_5}$. Because of the restricted domain of $r$ that we used to illustrate the behavior near ${\bf C_2}$, the figure doesn't show curves near the RSC approaching the saddle ${\bf C_3}$ and the RSC evolving from infinite $\Omav$ into ${\bf C_3}$, which is a sink for this worldline. 
%
%

%
%
%
%


\subsection{Re--collapsing wormhole}

In dust configurations without a RSC (``wormhole'' topology), the $\T$ can be homeomorphic to either $ \mathbb{S}^2\times \mathbb{R}$ or to $ \mathbb{S}^2\times \mathbb{S}^1$. These configurations were examined in reference~\cite{H87} (see also \cite{HM,HMM}) and can be constructed with $Y_i(r)$ having no zeroes in all its domain. Since any choice of such function will necessary fulfill $Y_i'(r^*)=0$ for at least a ``turning point'' $r=r^*$, these configurations can only be regular if they comply with the regularity condition (\ref{TPcond}), which implies a re--collapsing ``elliptic'' dynamics with $\Omiav-1>0$ (see Appendix A). 

In the integral definitions of $\rhoav$ and $\RRav$ in (\ref{rhoave}) and (\ref{3Rave}) we took the RSC as the lower bound of the integrals. This is a boundary condition on these definitions ensuring that regularity at that RSC is fulfilled~\cite{SG,HM,HMM}. If there are no RSC then this boundary condition can be specified by demanding that $m_i$ and $k_i$ vanish at an asymptotic value of $r$ along the $\T_i$. Such value can be then taken as the lower bound for the integrals. However, while this solves the mathematical problem, without a RSC the interpretation of $\rhoav$ and $\RRav$ as average functions becomes less clear. 

\begin{figure}[htbp]
\begin{center}
\includegraphics[width=2.5in]{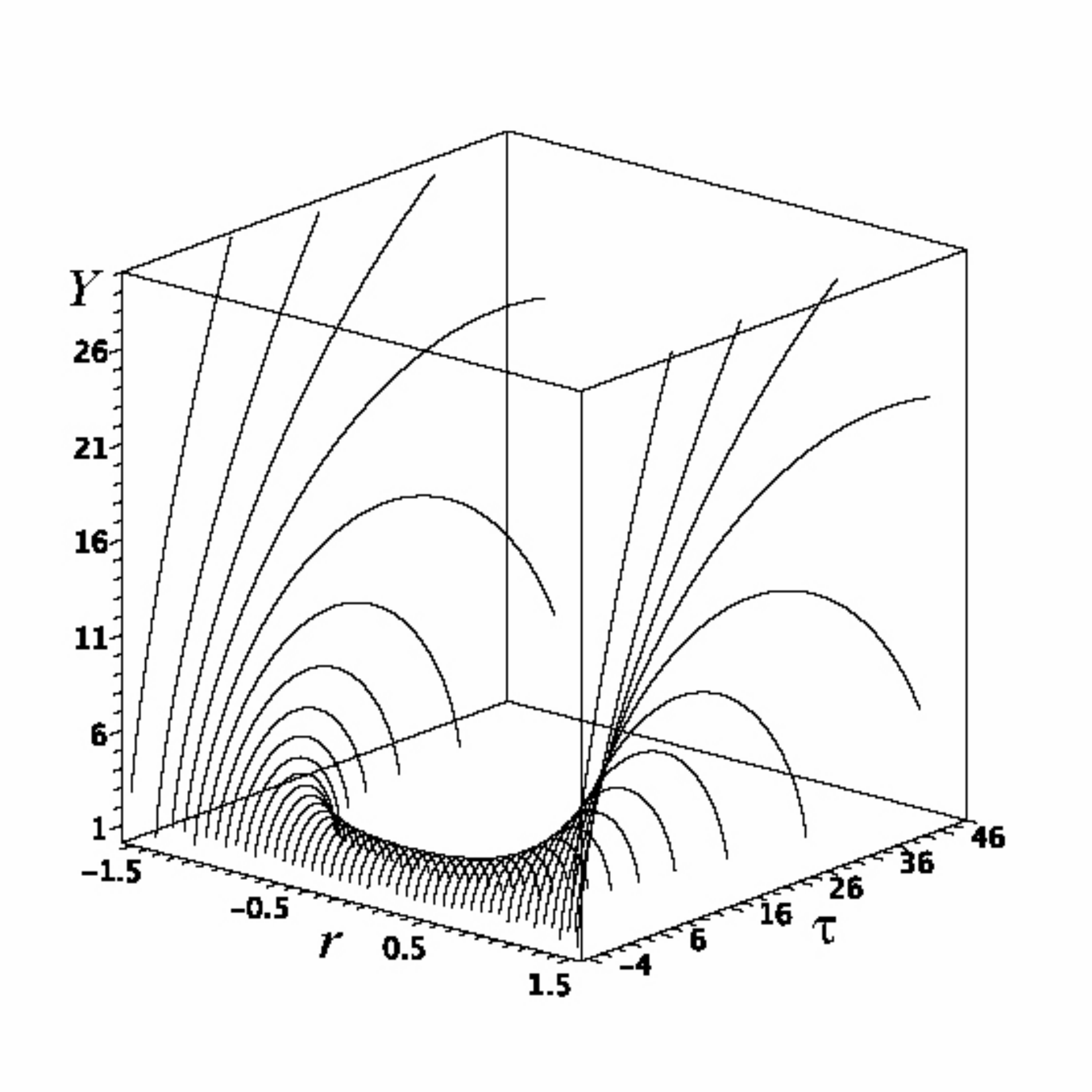}
\caption{{\bf Curvature radius of dust layers with wormhole topology.} Notice how for each hypersurface $\T$ with $\tau$ constant, $Y=Y_i\ell$ has no zeros in its regular range and takes its minimal value at the ``throat'' at $r=0$. Dust layers near the ``throat'' re--collapse in very a short interval of $\tau$, while layers further away re--collapse in much longer times. }
\label{wormhole2_Y}
\end{center}
\end{figure}

A choice of initial value functions for the re--collapsing wormhole is given by:
\ba m_i &=& \frac{m_{00}(1+\alpha_0)}{1+\alpha_0\,\sec^2 \,r},\qquad m_{00}=1.1,\quad \alpha_0=5.0\nonumber\\
    k_i &=&  \frac{k_{00}(1+\beta_0)}{1+\beta_0\,\sec^2\,r},\qquad k_{00}=0.48,\quad \beta_0=3.0\nonumber\\
    Y_i &=&  H_0^{-1}\,\sec\,r,\nonumber\\
    \label{wormhole_ivf}\ea
Notice that the ``turning point'' (or ``throat'') in this example is located at $r=r^*=0$ with (\ref{TPcond}) taking the form $\kiav(0)=1$. Also, both $m_i\to 0$ and $k_i\to 0$ as $r\to\pm \pi/2$. Therefore, we can define $\miav$ and $\kiav$ by integrals like (\ref{rhoave}) and (\ref{3Rave}) by taking the lower integration limit at $r=-\pi/2$ and integrating forward in $r$ up to $\pi/2$. The metric function $Y=Y_i\ell$ obtained from (\ref{sys3}) for this case is depicted in figure \ref{wormhole2_Y}.  
%
%

\begin{figure}[htbp]
\begin{center}
\includegraphics[width=2.5in]{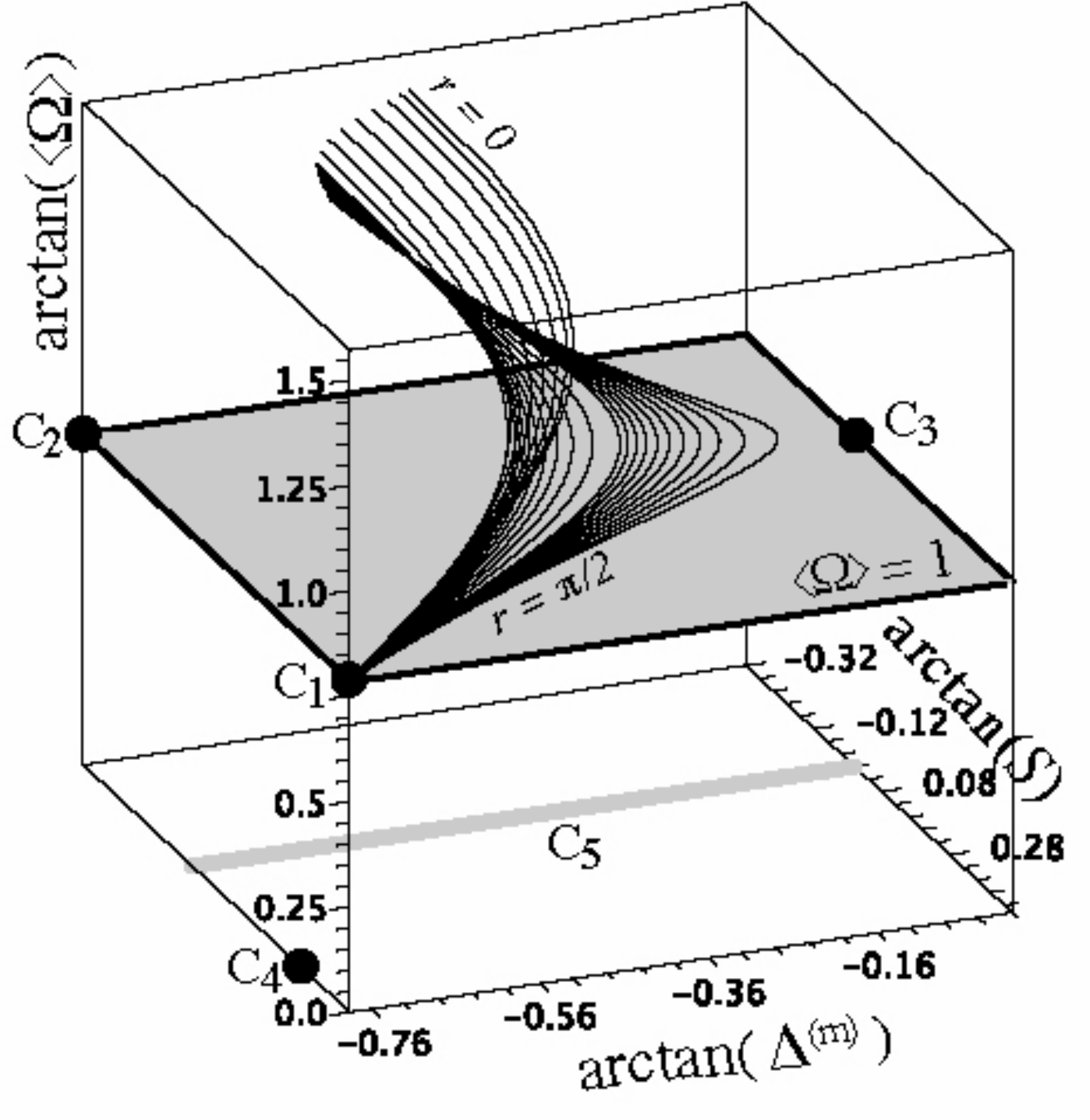}
\caption{{\bf Dust wormhole} There is no RSC. Notice how curves near the ``throat'' at $r=0$ rapidly reach maximal expansion with $\Omav$ diverging, while curves close to the asymptotic value $r=\pi/2$ take a long time to reach this stage and approach the saddle point ${\bf C_3}$ with $\Dm=S=0$. The lack of a RSC implies that ${\bf C_1}$ is now a global past attractor.}
\label{wormhole}
\end{center}
\end{figure}

The evolution of the wormhole configuration in phase space is depicted in figure \ref{wormhole}. Only curves in the range $0\leq r <\pi/2$ are displayed, since (because of the symmetric construction of the example) curves in the range $-\pi/2< r \leq 0$ behave identically. This evolution is similar to that of high density re--collapsing universes, with all solution curves starting at the past attractor or source ${\bf C_1}$ associated with an initial big bang singularity and rising towards maximal expansion $\Omav\to\infty$. The lack of a RSC is evident and this means that ${\bf C_1}$ is a global past attractor. 

It is interesting to notice how the ``throat'' and curves near it rapidly leap into diverging $\Omav$ because they reach maximal expansion and collapse very fast (see figure \ref{wormhole2_Y}), but curves close to $r=\pi/2$ spend a long time very near the plane $\Omav=1$ and approach the saddle ${\bf C_3}$ where $S=\Dm=0$.   
%
%
We took as example a very simple symmetric form of $\T_i$ with only one turning point, but it is possible to choose any number of such points~\cite{H87}. It is also possible to construct a configuration whose hypersurfaces $\T$ are homeomorphic to torii $ \mathbb{S}^2\times \mathbb{S}^1$ (see \cite{HM,HMM}).

\begin{figure}[htbp]
\begin{center}
\includegraphics[width=2.5in]{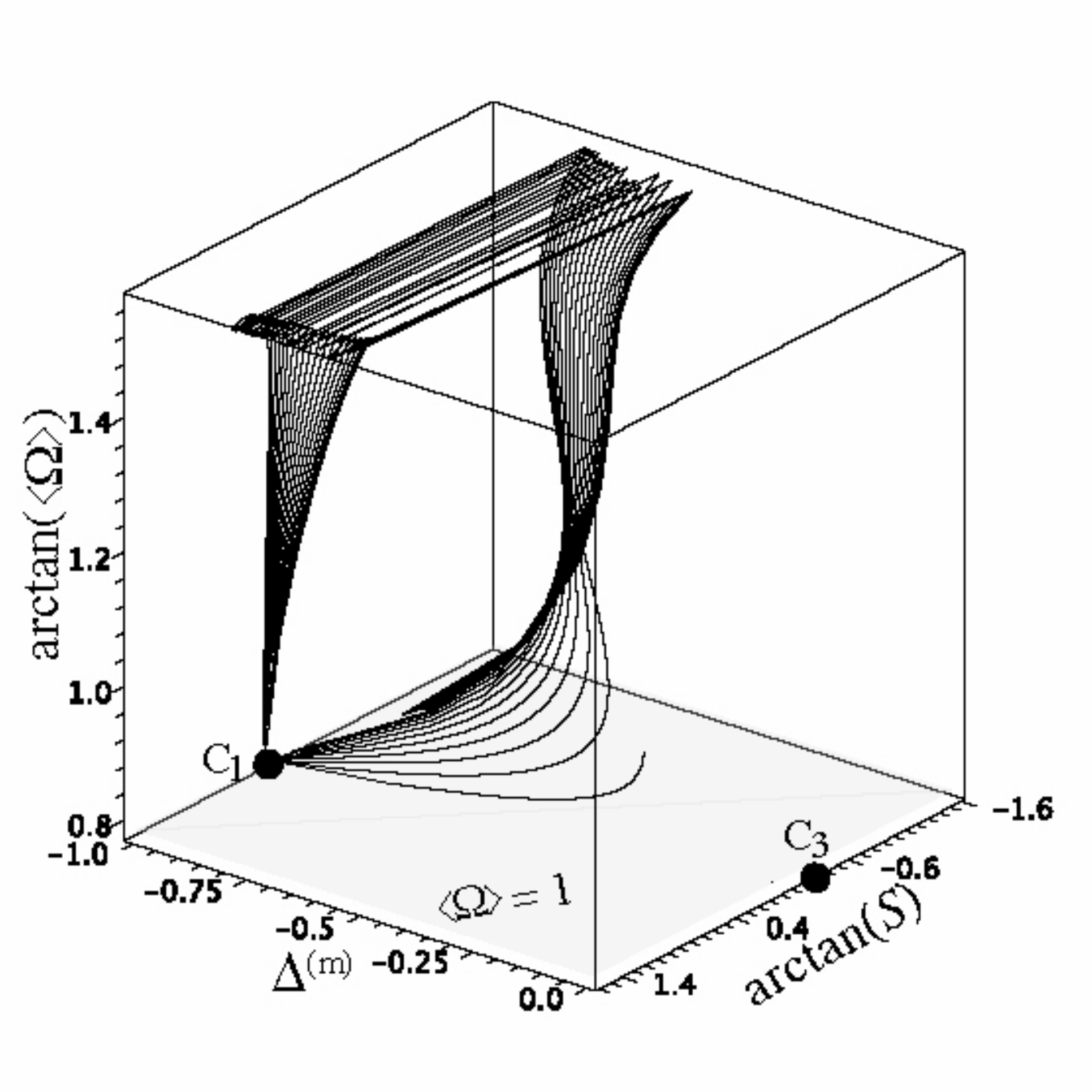}
\caption{{\bf Dust wormhole: collapsing stage} The figure describes the expanding and collapsing stages. The curves of figure \ref{wormhole} near the ``throat'' $r=0$ can be recognized and seen going upwards into infinite values of $\Omav$ and $S$, all in order to plunge downwards into ${\bf C_1}$. Notice that the lack of a RSC makes this critical point a global past and future attractor.}
\label{collapse_w}
\end{center}
\end{figure}

The collapsing stage for wormhole configurations is very similar to the stages depicted in figures \ref{collapse} and \ref{collapse_ct}. We show in figure \ref{collapse_w} the expanding and collapsing stage for curves near the ``throat'' $r=0$ of this configuration. It is easy to recognize some of the curves plotted in figure \ref{wormhole}, which continue towards infinite values of $\Omav$ and $S$, in order to plunge downwards to end in the critical point ${\bf C_1}$, which is a future attractor or sink in this stage.   
%
%
%


\subsection{Vacuum limit}

The evolution in phase space takes place in the plane $\Dm =-1$. As we commented in the previous section, for zero binding energy $\Omiav = 1$ (Lema\^\i tre coordinates \cite{Steph}) it goes from source ${\bf C_1}$ to sink ${\bf C_2}$, while for negative binding energy $\Omiav > 1$ (Novikov coordinates \cite{MTW}) it goes from source ${\bf C_1}$, approaches saddle ${\bf C_2}$ and goes upwards to $\Omav\to\infty$. The phase space evolution in this case is qualitatively analogous to that of a wormhole topology, which is not surprising since the Schwarzschild--Kruskal space--time has no RSC, hence ${\bf C_1}$ will be a global past and future attractor.  

\begin{figure}[htbp]
\begin{center}
\includegraphics[width=2.5in]{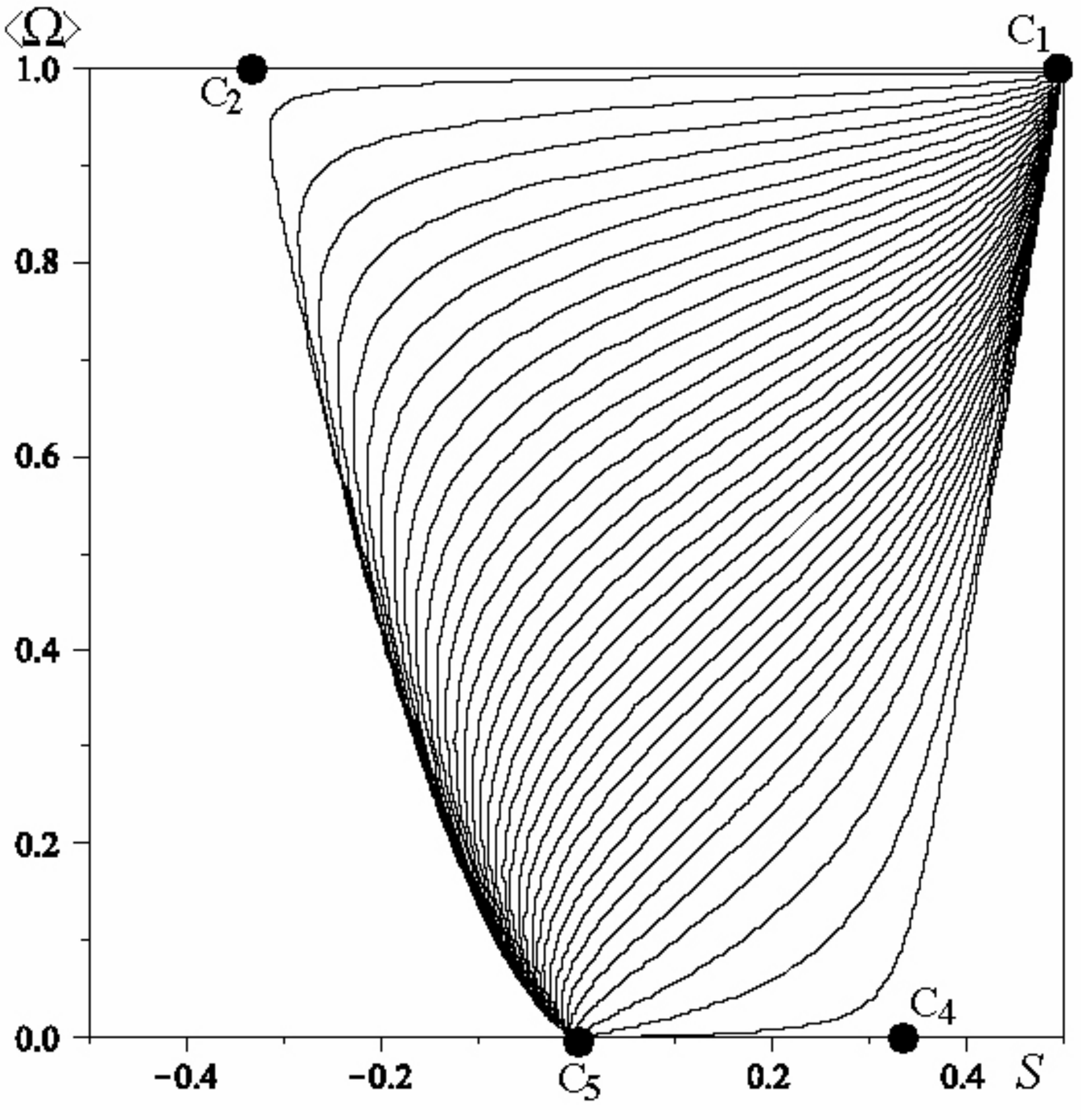}
\caption{{\bf Vacuum case}  The solution curves correspond to the Schwarzschild--Kruskal space--time in coordinates given by radial geodesic congruence that is perpetually expanding with positive binding energy (thus $\Omiav<1$). Solution curves are confined to the plane $\Dm = -1$. The curves evolve from the global past attractor or source ${\bf C_1}$ towards the global future attractor or  sink in ${\bf C_5}$ with $S=0$, approaching the saddles ${\bf C_2}$ and ${\bf C_4}$. }
\label{vacumcase}
\end{center}
\end{figure}

We provide here an example with positive binding energy $\Omiav < 1$.  As shown in figure \ref{vacumcase} the solution curves emerge from the global past attractor source ${\bf C_1}$, some curves approach the saddle ${\bf C_2}$ and some approach the saddle ${\bf C_4}$ and all terminate in the global future attractor or sink ${\bf C_5}$.

%
%

\section{Discussion and conclusion}

We have examined the class of dust LTB models in terms of the techniques generically known as ``dynamical systems'' as they evolve in a 3--dimensional phase space parametrized by the variables $[S,\Dm,\Omav]$, whose physical and geometric interpretation is intuitive and straightforward. We can say in general that if initial conditions are selected so that ``shell crossing'' singularities are avoided (condition (\ref{nxcr})), then the phase space variables $[S,\Dm]$ will be bounded. Also, the sign of $\Omiav(r_0)-1$ for any given solution curve $r=r_0$ determines if it will evolve for all $\xi$ in the domain of $\Omav(\xi,r_0)-1$ that has the same sign. The various invariant subspaces, particular cases and phase space evolution for representative dust configurations has been presented in detail in the previous sections. 

Except for the worldlines associated with RSC's, all ``off center'' solution curves begin at a past attractor given by a repelling source ${\bf C_1}$, characterized by $S=1/2$,\, $\Omav=1$ and $\Dm=-1$ and associated with a big bang initial singularity.   It is easy to show that ${\bf C_1}$ is indeed the only past attractor for all `off center'' solution curves: the asymptotic past regime for all curves is given by the limit $\xi\to\-\infty$ (or $\ell\to 0$), so from (\ref{Ome_ana}) we have $\Omav\to 1$ in this limit. On the other hand, we can write $\Dm$ and $\Dk$ as
\begin{equation}\Dm = \frac{m_i}{\miav\,\Gamma}-1,\qquad \Dk = \frac{k_i}{\kiav\,\Gamma}-1.\label{DmkG}\end{equation}
But, from the analytic forms given in Appendix C and for finite and regular $m_i$ and $k_i$, we have in the limit $\ell\to 0$ the following asymptotic behavior: \, $\Gamma \sim \ell^{-3/2}$ for $\Omiav=1$ \, and $\Gamma \sim \ell^{-1}$ for $\Omiav\ne 1$. Thus, from (\ref{DmkG}), we have $\Dm\to -1$ and $\Dk\to -1$ in this limit, but then, from (\ref{S_constr}) and bearing in mind that $\Omav\to 1$, we have $S\to 1/2$. 

As shown in figures \ref{collapse}, \ref{collapse_ct}, \ref{collapse_sf} and \ref{collapse_w}, all re--collapsing curves also terminate in ${\bf C_1}$, which is in this case a future attractor for these curves and is associated with the collapsing (``big crunch'') singularity. It is only the fact that RSC's do not evolve nor terminate (collapse) in ${\bf C_1}$ what prevents this attractor to be a global one, though in configurations lacking a RSC (wormhole topology and vacuum limits) this attractor is indeed global.

In order to prove that near ${\bf C_1}$ we have a self--similar regime we use the definitions (\ref{Sigma_}), (\ref{ell}), (\ref{H_def}) and (\ref{S_def}).  The condition $S= 1/2$ can be  reduced to
\ba Y_{\bf C_1}&=& r(1+\vartheta)^{2/3}, \qquad Y_{\bf C_1}' = \frac{1+\vartheta/3}{(1+\vartheta)^{1/3}},\nonumber\\
&\textrm{with}&\qquad \vartheta\equiv \frac{\tau}{r},\label{selfsim1} \ea
where $\tau=H_0 ct$ and we have eliminated two functions of the form $a(ct)$ and $b(r)$ by trivially re--labeling the time and radial coordinates. Since $\Omav=1$ implies $\kiav=0=\RRiav$, the metric (\ref{LTB1}) near ${\bf C_1}$ takes the form of the self--similar sub--case of LTB metrics~\cite{CC} 
\ba ds^2 &=& -d\tau^2 +A^2(\vartheta)\,dr^2+r^2B^2(\vartheta)\,(d\theta^2+\sin^2\theta\,d\phi^2),\nonumber\\  
&\textrm{with}&\quad A(\vartheta) = Y_{\bf C_1}',  \qquad r B(\vartheta) = Y_{\bf C_1}.\label{selfsim2}\ea
The fact that $\Dm=-1$ doesn't imply that ${\bf C_1}$ is a Schwarzschild vacuum, but that $\Gamma\to\infty$ as $\ell\to 0$. Evaluating $\Gamma$ for the self--similar metric forms (\ref{selfsim1}) and (\ref{selfsim2}) with $Y_i= Y_{\bf C_1}(t_i,r)$, we get
\begin{equation} \Gamma =\frac{3+\vartheta}{3+\vartheta_i}\,\frac{1+\vartheta_i}{1+\vartheta}\end{equation}
with $\vartheta_i=\tau_i/r$, which shows how $\Gamma\to\infty$ at the locus of the big bang singularity $\vartheta=-1$ where $Y_{\bf C_1}=r B(\vartheta)\to 0$. Notice that the self--similar case has no shell crossing singularity since $\Gamma$ vanishes at $\vartheta=-3$ but the evolution range is $\vartheta>-1$ (for collapsing stages we would have $Y$ and $Y'$ given by the same expressions as (\ref{selfsim1}), but with a minus sign, so that the ``big crunch'' happens before the locus of $\Gamma=0$). 

Therefore, we can consider the source ${\bf C_1}$ as marking a self--similar limit and since all ``off center'' solution curves start at this critical point (associated with a big bang singularity), we can say that dust LTB models have a self--similar behavior near this singularity. Since collapsing curves in re--collapsing configurations also end in ${\bf C_1}$ (a sink), we have self--similar behavior near the collapsing singularity.

Another interesting feature worth remarking is the role of the saddle ${\bf C_2}$ in the ``structure formation'' configurations discussed in section IX-D. As shown by figures \ref{struct_form3} and \ref{collapse_sf}, this saddle splits collapsing curves $\Omiav>1$ from the expanding ones $0<\Omiav<1$. The former evolve back into ${\bf C_1}$ (now a sink) while the latter fall into the line of sinks ${\bf C_5}$. This behavior denotes a basic instability of the invariant subspace of the ``parabolic'' evolution $\Omav =1$, since for any solution curve in this space any arbitrarily small perturbation $\delta\ll 1$ on initial conditions $\Omiav =1+\delta$ will trigger a radically different ``repulsive'' evolution away from $\Omav =1$, either to a collapsing regime ($\delta>0$) or to perpetual expansion ($\delta<0$). 

The splitting saddle ${\bf C_2}$ is marked by $\Dm=-1$,\,$S=-1/3$ and $\Omav=1$, it can be shown from the analytic solutions in Appendix C and from numerical tests that $\Gamma$ is finite as the curves approach these values ($\Gamma$ diverges for later times as the collapsing curves actually collapse). Therefore, the fact that $\Dm\to-1$ for curves approaching this saddle means that these curves experience near ${\bf C_2}$ unstable conditions similar to those of a Schwarzschild--Kruskal vacuum parametrized by Lema\^\i tre coordinates made with radial geodesics with zero binding energy ($\kav=0$ or $\Omav=1$). In fact, the point ${\bf C_2}$ represents the global future attractor (sink) for this marginally bound vacuum configuration.  

The association of the saddle ${\bf C_2}$ with a Schwarzschild--Kruskal vacuum roughly conforms with an intuitive picture. A very rough, hand waving, description of conditions near ${\bf C_2}$ could be:  collapsing dust layers ``retreat'' inwards towards a smaller collapsing region, while expanding layers ``retreat'' outwards in a fast expansion, thus we have a sort of unstable ``evacuation'' effect that is roughly similar to a Schwarzschild--Kruskal vacuum with the Schwarzschild mass being the accumulated effective mass $M_0=(1/2)\miav Y_i^3$ of the collapsing layers. It is known~\cite{HM,HMM} that in this type of configuration the more ``external'' layers in the collapsing region ($\Omiav>1$) take infinite time to collapse, so that the resulting black hole perpetually takes accretion from these ``border'' dust layers. However, we can assume that the mass contribution from this accretion is small and lengthy to come, so that once most ``inner'' layers with $\Omiav>1$ have collapsed the expanding observers close to $\Omiav=1$ do perceive a sort of approximate Schwarzschild--Kruskal vacuum with zero binding energy. 

We feel that the qualitative numerical treatment presented in this article can serve for studying applications of astrophysical interest of dust LTB solutions. Perhaps for this purpose the system (\ref{sys3}) can be more useful than the dynamical system (\ref{system5}). It is also interesting to see this methodology generically applied to generalizations of these solution, such as LTB--de Sitter (dust with cosmological constant) or the most general source compatible with the LTB metric: a mixture of dust and an inhomogeneous and anisotropic fluid, which can respectively model cold dark matter and dark energy. Extensions of this work along these lines are presently under consideration.      


%
\begin{acknowledgments}
The author acknowledges support from Instituto de F\'\i sica, Universidad de Guanajuato. 
\end{acknowledgments}

\section*{APPENDIX A: Regularity and singularities in dust LTB solutions}

An extensive discussion of regularity conditions and important geometric features of dust LTB solutions can be found in \cite{HM,HMM}. These features were also discussed with the variables of this article in \cite{SG}. We present here a brief review.

\subsection*{Regular Symmetry Centers (RSC) and topology of the $\T_i$}

LTB dust solutions admit up to two RSC's, which are the regular wordlines made up by the space--time evolution of the fixed point of the rotation group SO(3) where the orbits have zero area. In terms of the initial value functions, the RSC is the comoving worldline marked by a zero of the function $Y_i(r)$. Since the freedom to choose a radial coordinate in the LTB metric means that this function can be arbitrarily chosen, it is useful to select it as $Y_i=H_0^{-1}\,f(r)$, where $f(r)$ is a (at least) $C^2$ real function that describes in a simple manner the topological configuration or homeomorphic class for an initial $\T_i$. We have the following choices:
\bi 
\item ``Open topology'': $\T_i$ homeomorphic to $\mathbb{R}^3$,\

\begin{equation} f(r)=\tan r \quad 0\leq r<\pi/2\quad \textrm{One RSC at}\ r=0.\label{opentop} \end{equation}.

\item ``Closed topology'':  $\T_i$ homeomorphic to $\mathbb{S}^3$, \

\begin{equation} f(r)=\sin r \quad 0\leq r\leq\pi\quad \textrm{Two RSCs at}\ r=0,\,\pi. \label{closedtop} \end{equation}.

\item ``Wormhole'': $\T_i$ homeomorphic to $ \mathbb{S}^2\times \mathbb{R}$, \

\begin{equation} f(r)=\sec r \quad -\pi/2< r<\pi/2\quad \textrm{Zero RSCs}. \label{wormtop} \end{equation}.
\ei 
For the open topology $f(r)$ is a monotonously increasing function. In fact, we could have just selected $f=r$ for simplicity, but the choice $f=\tan r$ allows one to examine the asymptotic behavior associated with $Y_i\to\infty$ in solution curves of (\ref{system5}). 

In the wormhole topology the function $Y_i$ must be positive and without zeroes. Numerical tests show that monotonous functions without (at least) one ``turning value'' (zero of $Y_i'$) yield finite proper radial distances along the $\T_i$ as $r\to\pm\infty$ (depending on the choice of $Y_i$). This indicates that such functions are inadequate choices of the radial coordinate, as they do not provide a full analytic extension of the manifold. Therefore, we select $Y_i>0$ with at least a turning value $r=r^*$ within the open range of $r$. In the choice (\ref{wormtop}) this value is $r^*=0$ (the ``throat'' of the wormhole). 

In the ``closed'' topology we must have a turning value, which with the choice (\ref{closedtop}) is $r^*=\pi/2$. Since in both cases (closed and wormhole topologies) we have $Y_i'(r^*)=0$,  regularity of the metric function $g_{rr}$
$$\frac{Y'}{(1-K)^{1/2}}=\frac{\ell \,\Gamma\,Y_i'}{[1-H_0^2\langle k_i\rangle\,Y_i^2]^{1/2}}$$
requires (see \cite{SG,HM,HMM} and \cite{bon3}) that $\RRi$ be selected so that $K(r^*)=1$, or equivalently:
\begin{equation} \frac{1}{6}\RRiav(r^*) = H_0^2\,\kiav(r^*)= \frac{1}{Y_i^2(r^*)}.\label{TPcond} \end{equation}
Given a choice of homeomorphic class for the $\T_i$ all regular $\T$ have the same class (see \cite{SG}). All configurations examined numerically  corresponding to closed or wormhole topologies comply with (\ref{TPcond}).

\subsection*{Singularities} 

LTB dust solutions present two types of curvature singularities marked by $(ct,r)$ values such that:
\bse\label{sings}\ba \ell(ct,r) = 0,\quad \textrm{``bang'' or ``crush''},\\
\Gamma(ct,r) = 0, \quad \textrm{``shell crossing''}\ea\ese
The ``bang'' or ``crush'' singularities are an inherent feature and cannot be avoided, though it is always possible to select initial conditions so that ``shell crossing'' do not arise (either for all the evolution time or for a given range $t\geq t_i$). 

Initial conditions to avoid unphysical shell crossing singularities were given in terms of the original LTB variables by Hellaby and Lake~\cite{HL} (see \cite{HM,HMM}), and in terms of the initial value functions defined in this article by \cite{SG}. For broadly generic configurations that exclude very special (and unstable) features, we have:
\bse\label{nxcr_ini}\ba &\textrm{for}&\ 0<\Omiav\leq 1:\nonumber\\
&{}& \quad -1<\Dim\leq 0, \quad -2/3 < \Dik < 2/3,\label{nxcr_ini_1}\\
&{}&\nonumber\\
&\textrm{for}&\ \Omiav> 1:\nonumber\\
&{}& \quad -1<\Dim\leq 0, \quad -2/3<\Dik\leq 0,\nonumber\\ 
&{}&\quad 2\pi P_i\left[\Dim-\frac{3}{2}\Dik\right]\geq \left[P_iQ_i-1\right]\Dim\nonumber\\
&{}&\quad+\left[\frac{3}{2}P_iQ_i-1\right]\Dik\geq0,,\label{nxcr_ini_2}
\ea\ese
where
\bse\ba P_i &=& \frac{\sqrt{2-x}}{x^{3/2}},\nonumber\\
 Q_i &=& \arccos (1-x)-\sqrt{x(2-x)},\nonumber\\
x &\equiv& \frac{\RRiav}{\kappa c^2\rho_i}=\frac{2\langle k_i\rangle}{\langle m_i\rangle}=\frac{2[\Omiav-1]}{\Omiav}\nonumber.\ea\ese

Equations (\ref{nxcr_ini}) clearly show how initial conditions characterized by density and 3--curvature clumps (see (\ref{Dm_CV})) lead generically to avoidance of a shell crossing singularity if $0<\Omiav\leq 1$, and as a corollary, to finite and bounded values of $\Dm$ and $S$ along solution curves. This regularity is more stringent for the case $\Omiav >1$. However, even if (\ref{nxcr_ini}) hold at the $\T_i$ it is still possible to have density or 3--curvature voids forming along the evolution of the curves, all without violating (\ref{nxcr}). This was discussed by~\cite{mustapha}, but their methodology is too complicated. It is more clearly seen here from the fact that the scaling laws (\ref{DmSL}) and (\ref{DkSL}) do not, necessarily, forbid $\Dm$ and/or $\Dk$ from reversing the sign of the initial $\Dim$ and $\Dik$. The configuration discussed in section IX--B is a numerical example of this type of ``clump turning into void'' configuration.   

\section*{APPENDIX B: An equivalent dynamical system}

We have examined the evolution of solution curves of the autonomous system (\ref{system5}) in the 3--dimensional phase space parametrized by $[S,\,\Dm,\,\Omav]$. However, even if (\ref{system5}) has only derivatives with respect to the evolution parameter $\xi$, the differential equations are still PDE's. We provide below formal proof that such a system of autonomous evolution PDE's is equivalent to a system of ODE's with initial conditions restricted by the fulfillment of space-like constraints.

Let $\FF_{(r)} \subset \mathbb{R}$ and $\FF_{(\xi;r)} \subset \mathbb{R}$ be,
respectively, the regularity range of $r$ and the maximal range of extendibility of
$\xi$ for a given $r$, the system (\ref{system5}) can be associated with the flow
$\Phi_\xi:\mathbb{R}^3\to\mathbb{R}^3$ of the vector field $\partial/\partial\xi$ given by
\begin{equation}\Phi_\xi(\vec X_i) = \vec X,\label{flowPDE}\end{equation}
where $\vec X_i:\FF_{(r)}\to \mathbb{R}^3$ is the initial state represented as a curve in
$\mathbb{R}^3$ given parametrically as
\begin{equation} \vec X_i(r)=[S_i(r),\,\Dim(r),\,\Omiavr],\label{init_st}\end{equation}
and  $\vec X: \FF_{(\xi;r)}\times \FF_{(r)}\to \mathbb{R}^3$ is a solution
surface associated with $\vec X_i$ by means of (\ref{flowPDE}). This surface can be
represented parametrically as the following surface in
$\mathbb{R}^3$:
\begin{equation} \vec
X(\xi,r)=[S(\xi,r),\,\Dm(\xi,r),\,\Omavxr],\label{sol_surf}\end{equation}
so that $\vec X_i(r)=\vec X(0,r)$. Given a point $\vec X_i(r_0) \in
\mathbb{R}^3$ there is a unique orbit or solution curve $\C_{r_0}: \FF_{(\xi;r_0)}\to
\mathbb{R}^3$ of (\ref{system5}) which can be represented parametrically as:
\begin{equation}\C_{r_0}(\xi) = \vec X(\xi,r_0),\label{sol_curve}\end{equation}
so that $\C_{r_0}(0) =\vec X_i(r_0)$. The set of solution curves $\C_{r}(\xi)$ for all
$r\in \FF_{(r)}$ generates the solution surface $\vec X(\xi,r)$.

In order to relate (\ref{system5}) with a system of ordinary differential equations, we
notice that every solution curve (\ref{sol_curve}) associated with a fixed $r=r_0$ is a
solution of the following system:
\bse\label{system6}\ba \frac{d S_0}{d \xi} &=&
S_0\left(3\,S_0-1\right)+\frac{1}{2}\left(\Dm_0+S_0\right)\Omav_0,\nonumber\\
\label{ev_S6}
\\
\frac{d \Dm_0}{d \xi} &=& 3\,S_0\,\left[\,1+ \Dm_0 \right],\label{ev_Dm6}
\\
\frac{d \Omav_0}{d \xi} &=&
\Omav_0\,\left[\Omav_0-1\right],\label{ev_Omega6}\ea\ese
where the functions $(S_0,\,\Dm_0,\,\Omav_0) : \FF_{(\xi;r_0)}\to \mathbb{R}$ are given by
\bse\ba S_0 &=& S(\xi,r_0),\\
\Dm_0 &=& \Dm(\xi,r_0),\\
\Omav_0 &=& \langle\Omega(\xi,r_0)\rangle,\ea\ese
and comply with the initial conditions:
\begin{equation} [S_0(0),\,\Dm_0(0),\,\Omav_0(0)]=\vec
X_i(r_0).\label{init_st0}\end{equation}
The fact that every solution curve of (\ref{system5}) will be a solution of a system like
(\ref{system6}) for initial conditions given as in (\ref{init_st0}), suggests
considering a system of ordinary differential equations associated with the same flow
as (\ref{flowPDE})      
\bse\label{system7}\ba \frac{d s}{d \xi} &=&
s\left(3\,s-1\right)+\frac{1}{2}\left(\delta+s\right)\,\omega,\label{ev_S7}
\\
\frac{d \delta}{d \xi} &=& 3\,s\,\left[\,1+ \delta \right],\label{ev_Dm7}
\\
\frac{d \omega}{d \xi} &=&
\omega\,\left[\omega-1\right],\label{ev_Omega7}\ea\ese
for an initial state:
\begin{equation}\vec x_0 =
[s(0),\delta(0),\omega(0)]=[s_0,\delta_0,\omega_0],\label{init_st00}\end{equation}
and the solution surface given by a parametric form similar to
(\ref{sol_surf}): 
\begin{equation}\vec x =
[s(\xi,s_0),\,\delta(\xi,\delta_0),\,\omega(\xi,\omega_0)],\label{sol_surf00}\end{equation} 
so that $\vec x_0=\vec x(0)$. Since the solution surface of this system can depend
smoothly on initial conditions $\vec x_0$,  each one of the functions $(s,\delta,\omega)$
solving (\ref{system7}) is a one--parameter family of functions. In principle, there are
many ways in which the initial state (\ref{init_st00}) can be parametrized. In particular,
this state can be given as any curve
$[s_0(\alpha),\delta_0(\alpha),\omega_0(\alpha)]$ in $\mathbb{R}^3$ that intersects the
solution curves $\vec x(\xi,\vec x_0)$ only in one point, hence we can always find
suitable one--parameter functions $s_0(\alpha),\delta_0(\alpha),\omega_0(\alpha) $ such
that $\vec x$ is expressible as:
\begin{equation}\vec x(\xi,\alpha) =
[s(\xi,s_0(\alpha)),\,\delta(\xi,\delta_0(\alpha)),\,\omega(\xi,\omega_0(\alpha))].
\label{sol_surf11}
\end{equation} 

If, in particular, the initial state (\ref{init_st00}) is parametrized as:
\begin{equation}\vec
x_0(\alpha)=[s_0(\alpha),\delta_0(\alpha),\omega_0(\alpha)]=\vec
X_i(\alpha),\label{init_constr}\end{equation}  
with $\alpha \in \FF_{(r)}$, then for every $\alpha$ there is a $C^\infty$ inclusion map
$\psi:\mathbb{R}^3\to\mathbb{R}^3$ mapping every solution curve of (\ref{system5})
given by (\ref{sol_curve}) for a fixed $r\in\FF_{(r)}$ to a solution curve of the system
(\ref{system7}) given by (\ref{sol_surf11}) with a fixed $\alpha=r$. That is
\begin{equation}\psi(\vec X(\xi,r)) = \vec x(\xi,\alpha),\label{incl_map} \end{equation}
and so, the solution surfaces of (\ref{system5}) are subsets of the solution surfaces of
(\ref{system7}) obtained by restricting the initial states by the constraint
(\ref{init_constr}). 

Notice that (\ref{system7}) admits many solution curves that do not comply with
(\ref{init_constr}), and so are incompatible with the constraints 
(\ref{initfuns})--(\ref{Si}) and so they are incompatible with the
evolution of an inhomogeneous dust source (LTB solution) complying with Einstein's field equations
expressed in terms of the 3+1 decomposition~\cite{EVE} under the
regularity assumptions that we have made (see Appendix A).

However, the solution curves of (\ref{system7}) satisfying (\ref{init_constr}) are
equivalent (under (\ref{incl_map})) to solution curves of (\ref{system5}) associated with
the flow (\ref{flowPDE}). Since (\ref{system7}) is a system of autonomous ODE's, geometric
features such as critical points and invariant spaces are well defined, so the
qualitative analysis of (\ref{system5}) can be performed on (\ref{system7}) as long as we
only consider its solution surfaces satisfying (\ref{init_constr}). In other words, we can
say that the dynamical system associated to inhomogeneous dust LTB solutions is a sort of ``reduced'' version of (\ref{system7}) with its
initial states restricted by (\ref{init_constr}). Since, as we have shown in sections IV and V, the space--like constraints are satisfied at all space--like slices once they hold at the initial $\T_i$, the incorporation of the radial
dependence of the solutions only requires this restriction on the initial states.  

In order to simplify the notation we found it useful to keep the same names for the variables
$S,\Dm,\Omav$ and to keep referring to (\ref{system5}),  with the understanding that the system we are really considering is (\ref{system7}),
thus every mention of solution curves and surfaces of (\ref{system5}) refers to solution curves and surfaces of (\ref{system7}) mapped by (\ref{incl_map})
and complying with (\ref{init_constr}).
 
\section*{Appendix C: Analytic solutions}

Analytic solutions for LTB dust solutions follow from solving the evolution equation (\ref{Ysq_dust}):
\begin{equation} \dot Y^2 = \frac{2M}{Y}-K.\nonumber\end{equation}
Using the variables defined in sections III and IV, this equation becomes equation (\ref{ellsq_dust}):
\begin{equation}\left[\frac{\partial \ell}{\partial\tau}\right]^2=\HH_i^2\left[\frac{\Omiav}{\ell}-(\Omiav-1)\right].\nonumber\end{equation}
where:
\ba \ell &=& \frac{Y}{Y_i},\quad \tau=H_0\,ct\nonumber
\\ 
2M &=& \miav\,H_0^2\,Y_i^3,\quad K=\kiav\,H_0^2\,Y_i^2,\nonumber
\\
\Omiav &=& \frac{\miav}{\HH_i^2}=\frac{\miav}{\miav-\kiav},\nonumber
\\
\Omiav -1 &=&\frac{\kiav}{\HH_i^2}=\frac{\kiav}{\miav-\kiav},\nonumber\ea
We summarize the solutions of this equation below

\subsection*{Parabolic solutions}

We have: $\Omiav=1$, so that $\kiav=0$ and $\HH_i^2=\miav$. These are the
only ones that can be given explicitly as
$\ell=\ell(\tau,r)$ from:
\begin{equation}  \tau = \tau_i \pm
\frac{2\,(\ell^{3/2}-1)}{3\HH_i},\label{parab2}\end{equation}
where the $\pm$ sign describes expanding ($+$) and collapsing ($-$) dust layers,
respectively characterized by $\tau>\taub,\,\,\ell_{,\tau} >0$ and
$\tau<\taub,\,\,\ell_{,\tau} <0$. Notice that the ``big bang time''
$\taub=H_0 c\tbb$ is no longer an independent function but must be found by setting
$\ell=0$ in (\ref{parab2}):
\begin{equation} \taub=\tau_i \mp
\frac{2}{3\HH_i},\end{equation} 
where now expanding and collapsing layers respectively correspond to ``$-$'' and ``$+$''. 

\subsection*{Hyperbolic solutions}

We have: $0<\Omiav<1$, so that $\kiav<0$ and $\HH_i^2=\miav+|\kiav|$. In this case we cannot obtain $\ell=\ell(\tau,r)$ in closed explicit form, but in parametric
form: $[\ell(\eta,r),\,\tau(\eta,r)]$ or implicitly as $\tau=\tau(\ell,r)$. The solutions are given in parametric form as:
\bse\label{hyppar2}\ba \ell(\eta,r) &=&
\frac{\Omiav\,\left[\cosh\,\eta -1\right]}{2\,(1-\Omiav)},\label{hyppar21}\\
\tau(\eta,r) &=& \tau_i \pm
\frac{\Omiav\,[(\sinh\eta-\eta)-(\sinh\eta_i-\eta_i)]}{2\HH_i\,(1-\Omiav)^{3/2}},
\nonumber\\
\label{hyppar22}\ea\ese
where  
\ba\eta_i =\textrm{arccosh}\left(\frac{2}{\Omiav}-1\right),\quad
\sinh\,\eta_i=\frac{2\sqrt{1-\Omiav}}{\Omiav}.\label{hyppar23}\ea
and $\pm$ denote expanding (+) and collapsing (--) layers. 
The coordinate locus of the central singularity (the ``big bang time'' function $\taub$) can be obtained by setting $\eta=0$ in (\ref{hyppar22}), ( so that $\ell=0$) leading to 
\begin{equation} \taub=\tau_i \mp
\frac{\Omiav\,(\sinh\eta_i-\eta_i)}{2\HH_i\,(1-\Omiav)^{3/2}},
\label{hypbangtime}\end{equation}
where now ``--'' and ``+'' respectively correspond to expanding and collapsing
layers. 

\subsection*{Elliptic solutions}

We have: $\Omiav>1$, so that $\kiav>0$ and $\HH_i^2=\miav-\kiav$. The parametric solutions are
\bse\label{ellpar2}\ba \ell(\eta,r) &=&
\frac{\Omiav\,\left[1-\cos\,\eta \right]}{2(\Omiav-1)},\label{ellpar21}\\
\tau(\eta,r) &=& \tau_i +
\frac{\Omiav\,[(\eta-\sin\eta)-(\eta_i-\sin\eta_i)]}{2\HH_i\,(\Omiav-1)^{3/2}},
\nonumber\\
\label{ellpar22}\ea\ese
where
\ba\eta_i =\textrm{arccos}\left(\frac{2}{\Omiav}-1\right),\quad
\sin\,\eta_i=\frac{2\sqrt{\Omiav-1}}{\Omiav}.\label{ellpar23}\ea
As with the hyperbolic case, the locus of the central singularity (the ``big bang time'', $\taub$), follows by setting $\eta=0$ in 
(\ref{ellpar22}), we get
\begin{equation} \taub= \tau_i -
\frac{\Omiav\,(\eta_i-\sin\eta_i)}{2\HH_i\,(\Omiav-1)^{3/2}},
\label{ellbangtime}\end{equation}
Dust layers in elliptic solutions reach a maximal expansion as $\ell_{,\tau}=0$ in (\ref{ellsq_dust}), corresponding to 
\bse\label{ellmax}\ba\ell = \ell_{_{\textrm{max}}}=\frac{\Omiav}{\Omiav-1},\qquad \eta =
\pi,\\
=\taub + \frac{\pi\,\Omiav}{2\,\HH_i\,(\Omiav-1)^{3/2}},\label{lmax}\ea\ese

For $\ell_{,\tau}<0$ dust layers collapse and so a collapsing phase is given by 
parametric solution (\ref{ellpar22}) for $\pi<\eta<2\pi$). The coordinate locus of the collapsing singularity (``big crunch''), $\tau_{_{bc}}$,
follows from setting $\eta=2\pi$ in (\ref{ellpar22}):
\begin{equation}\tau_{_{bc}}=\taub + \frac{\pi\,
\Omiav}{\HH_i\,(\Omiav-1)^{3/2}}.\label{ellcrunchtime}\end{equation} 

Since the expanding and collapsing singularities are respectively marked by $\eta=0$ and $\eta=2\pi$, then the layers' evolution is symmetric with respect to $\eta=\pi$ marking
the maximal expansion (\ref{ellmax}). However, as shown by (\ref{ellbangtime}),
(\ref{ellmax}) and (\ref{ellcrunchtime}), the maximal expansion time depends on the IVF's
$\Omiav$ and $\HH_i$, so it does not coincide (in general) with any $\T$ (hypersurface of
constant $\tau$) and the evolution for general IVF's $\Omiav,\,\HH_i$ is not
``time-symmetric'' with respect to the maximal expansion. 

While for the hyperbolic and elliptic cases we do not have an explicit closed form
solution $\ell=\ell(\tau,r)$, it is possible to use the parametric solutions
presented so far to visualize graphically the function $\ell(\tau,r)$. This can be done by
plotting the 3--dimensional parametric surfaces:
\begin{equation}[\tau(\eta,r), r, \ell(\eta,r)],\label{ell_parsur}\end{equation}
where the radial dependence is completely specified by the IVF's that enter in
(\ref{hyppar2})--(\ref{hyppar23}) and (\ref{ellpar2})--(\ref{ellpar23}).

\end{document}